\documentclass[useAMS, usegraphicx,usenatbib]{mn2e}
\usepackage[utf8]{inputenc}
\usepackage[T1]{fontenc}
\usepackage{graphicx}
\usepackage[fleqn]{amsmath}
\usepackage[figuresright]{rotating}
\usepackage{booktabs}
\usepackage{mathrsfs}
\usepackage{amssymb}
\usepackage[english]{babel}
\usepackage{amsfonts}
\usepackage{fontenc}
\usepackage{txfonts}
\usepackage{textcomp}
\usepackage{microtype}
\usepackage{aas_macros}
\usepackage[pdftex]{lscape}
\usepackage[usenames,dvipsnames]{xcolor}
\usepackage{lastpage}
\usepackage{subfig}
\usepackage[pdftex]{hyperref}

\bibpunct{(}{)}{;}{a}{}{,}

\newcommand*{\Lx}{L_{\rm X}}
\newcommand*{\Tx}{T_{\rm X}}
\newcommand*{\ephi}{{\hat{{\boldsymbol{e}}}}_{\varphi}}
\newcommand*{\vvphi}{v_{\varphi}}
\newcommand*{\dtpartial}[1]{\dfrac{\upartial#1}{\upartial t}}
\newcommand*{\diver}{\nabla\cdot}
\newcommand*{\convective}[1]{\left(#1 \cdot \nabla \right)}
\newcommand*{\grad}{\nabla}
\newcommand*{\norma}[1]{\left\lVert#1 \right\rVert}
\newcommand*{\Tr}{\mathrm{Tr}}
\newcommand*{\diff}{\mathop{}\!\mathrm{d}}
\newcommand*{\referee}[1]{#1}

\title[X-ray haloes and star formation]
{X-ray haloes and star formation in early-type galaxies}
 \author[A. Negri, S. Pellegrini, \& L. Ciotti]{Andrea Negri$^{1,2}$\thanks{E-mail:
andrea.negri@unibo.it}, Silvia Pellegrini$^1$ \& Luca Ciotti$^1$ 
\\$^1$ Department of Physics and Astronomy, University of Bologna, viale Berti Pichat
6/2, I-40127 Bologna, Italy\\$^2$ Institut d'Astrophysique de Paris, 98bis Boulevard Arago, F-75014 Paris, France}

\date{Accepted 2015 April 28. Submitted 2015 January 28.}

\pagerange{\pageref{firstpage}--\pageref{LastPage}} \pubyear{2015}

\begin{document}
\maketitle
\label{firstpage}
% \graphicspath{{tmp/}}

\begin{abstract}
  High resolution 2D hydrodynamical simulations describing the
  evolution of the hot ISM in axisymmetric
  two-component models of early-type galaxies well reproduced the observed trends 
of the X-ray luminosity ($\Lx$) and temperature ($\Tx$) with galaxy shape and rotation, however
they also revealed the formation
  of an exceedingly  massive cooled gas disc in rotating systems.  In a
  follow-up of this study, here we investigate the effects  of
  star formation in the disc, including the consequent injection of mass, momentum
and energy in the pre-existing interstellar medium. 
 It is found that subsequent generations of stars originate one after the other in the  
equatorial region; the mean age of the new stars is $> 5$ Gyr, and 
the adopted recipe for star formation can reproduce the empirical Kennicutt-Schmidt \referee{relation}.
The results of the previous investigation without star formation, 
concerning $\Lx$ and $\Tx$ of the hot gas, and their trends with galactic shape and rotation, are confirmed. 
At the same time, the consumption of most of the
cold gas disc into new stars leads to more realistic final
systems, whose cold gas mass and  star formation rate agree well with those observed in the local universe.
In particular, our models could explain the observation of kinematically aligned gas in
massive, fast-rotating early-type galaxies.
\end{abstract}


\section{Introduction} 

Early-Type galaxies (ETGs) are embedded in hot gaseous haloes produced
mainly by stellar winds, and heated to X-ray temperatures by Type Ia
supernovae (SNIa) explosions and by the thermalization of stellar
motions \citep{fabbiano1989, o'sullivan2001, ciottietal1991,
david.etal1991}. The thermalization is due to the interaction
between the stellar and SNIa ejecta and the pre-existing hot ISM
\citep[e.g.,][]{mathews1989, parriott.bregman2008}. Recent high resolution
2D hydrodynamical simulations of hot gas flows (\citealt{negri.etal2014}; \citealt{negri.etal2014b}, hereafter N14) showed that the presence of ordered
rotation in
the stellar component can alter significantly the ISM evolution with
respect to that shown by fully velocity dispersion supported systems of
same total mass and mass distribution.
Firstly, it is found that the rotation field of the ISM in rotating galaxies is very
similar to that of the stars, with a consequent negligible heating
contribution from thermalization of the ordered motions.  Secondly,
conservation of angular momentum in the ISM of rotating galaxies
results in the formation of a centrifugally supported cold equatorial disc, with the consequent reduction
of both the X-ray luminosity $\Lx$ and temperature $\Tx$ of the hot
ISM.  These results compared well with observations, which show a
dependence of $\Lx$ and $\Tx$ on the galactic shape and internal
dynamics: $\Lx$ is observed to be high only in round and slowly
rotating galaxies, and is limited to lower values for flatter, fast
rotating ones \citep{eskridge.etal1995, pellegrini.etal1997, sarzi.etal2010,
pellegrini2012, li.etal2011b, sarzi.etal.2013}; $\Tx$ of slowly
rotating systems is consistent just with the thermalization of the
stellar random kinetic energy, estimated from $\sigma_{\mathrm e}$ (the stellar velocity dispersion averaged within one effective radius $R_{\rm e}$),
while fast rotating systems show $\Tx$ values below 0.4~keV, and not
scaling with $\sigma _{\mathrm e}$ (\citealt{sarzi.etal.2013};
see also \citealt{pellegrini2011, posacki.etal2013}).

A major outcome of the previous simulations is that cold material can
be accumulated in considerable amounts, during the lifetime of rotating
galaxies. The cold gas typically settles in the equatorial plane,
where it forms an extended disc (of 0.5--10 kpc radius), that can be as
massive as $\simeq 10^{10}$ M$_{\sun}$, in the largest galaxies.  This
result brings in a few important questions: are these cold discs
observed? do they become, as seems natural, a site for star formation
(hereafter SF)?  is SF observed? and what is the impact of SF in the disc on $\Lx$ and $\Tx$?  
In recent years evidence
has been accumulating that ETGs host significant quantities of cold
gas, in the form of atomic and molecular hydrogen
\citep{morganti.etal2006, combes.etal2007, diseregoalighieri2007,
grossi.etal2009, young.etal2011}; approximately 50\% of massive ETGs
(of stellar mass $M_*\ga 10^{10}$M$_{\sun}$) contain $10^7$ to $10^9$
M$_{\sun}$ of HI and/or H$_2$ \citep{young.etal2014, serra.etal2012,
serra.etal2014, davis.etal2011, davis.etal2013}. A large, systematic
investigation of
the ATLAS$^\mathrm{3D}$ sample of ETGs with the Westerbork Synthesis Radio
Telescope found that $\simeq 40$\% of galaxies outside Virgo, and $\simeq
10$\% of galaxies inside it, are detected in HI, with $M_{\rm HI}\ga
10^7$M$_{\sun}$; the majority (2/3) of the detections consists of
settled configurations, where the cold gas is in discs or rings.  Small
discs (size of a few kpc), confined within the stellar body, share the
same kinematics of the stars; large discs (up to $5\times 10^9$
M$_{\sun}$) extend to tens of kpc, and in half of the cases are
kinematically misaligned with the stars \citep{serra.etal2012}. In
particular, fast-rotating Virgo galaxies have kinematically aligned
gas, and the most massive ($M_*\ga 8\times 10^{10}$M$_{\sun}$)
fast-rotating ETGs always have kinematically aligned gas, independent
of environment; this alignment leads to hypothesize that the gas has
been internally generated \citep{davis.etal2011, davis.etal2013}.  The HI
discs/rings around slowly-rotating ETGs, instead, are usually not
fully settled, which suggests an external origin (in mergers, or in
accretion from satellite galaxies, or from the intergalactic medium);
an external origin was considered likely also for those ETGs showing
stellar/gas kinematic misalignment.  Interestingly, while HI is more
ubiquitous, molecular gas is detected only in fast rotators across the
entire ATLAS$^{\rm 3D}$ sample \citep{young.etal2011, young.etal2014,
davis.etal2011, davis.etal2013}.

Besides giving indication about its origin through its morphology and
kinematics, the observed cold gas seems also to provide material for
SF.  Low level SF activity is present in $\sim $ one-third of ETGs
\citep{yi.etal2005, suh.etal2010, ko.etal2014}; \citet{sarzi.etal2006}
showed ongoing SF signatures in the optical spectra of at least $\sim
10$\% of nearby ETGs.  In the ATLAS$^\mathrm{3D}$ sample, galaxies with HI
within $\sim$ 1 $R_\mathrm{e}$  exhibit ongoing SF in $\sim 70$\% of
the cases, $\sim 5$ times more frequently than galaxies without HI
%ETGs with small gas discs seem to convert HI into H$_2$ as efficiently as spirals 
\citep{serra.etal2012}.  Interestingly, as for molecular gas,
some degree of SF and young stellar populations are detected only in
fast rotators, in the ATLAS$^\mathrm{3D}$ sample \citep{kuntschner.etal2010,
sarzi.etal.2013}. Integral-field spectroscopy showed that ETGs host
frequently a rotating stellar component younger and more metal rich
than the bulge \citep{krajnovic.etal2008}. 
The presence of this component, and the occurrence of SF, imply an important role for the cold gas during the evolution of ETGs \citep{khochfar.etal2011, naab.etal2014, cappellari.etal2013}. 

Finally, the cold gas content of ETGs is also an
important prediction of $\Lambda$CDM hydrodynamical simulations of
galaxy formation, that include cold gas evolution
\citep[e.g.,][]{martig.etal2013}, and 
accretion via various processes during the secular evolution of galaxies
(\citealt{oser.etal2010}; see also \citealt{lagos.etal2014};
\citealt{dubois.etal2013}).

In conclusion, the cold gas has become a tool to gain
insight into recent (and less recent) galaxy evolution.  In order to
correctly interpret the variety of observational results, and to use
them properly as constraints for different scenarios for the origin of
the structure and SF history of ETGs, it is crucial to
establish what is the relative importance of the various gas
production processes (internal and external to galaxies), gas
depletion ones (AGN and SF-driven outflows, environmental stripping),
and gas consumption ones (SF).  Cold gas is usually thought to come
from accretion from the surrounding medium or satellites, as well as
from gas-rich mergers; in the cases of the giant central-dominant
galaxies in groups or clusters it can also come from cooling of hot
gas \citep{edge.etal2010, mcdonald.etal2011}. The numerical
investigation of N14 showed an additional {\it internal } contribution
of cold gas, coming from the evolution of the passive stellar
population, that can be substantial, or even too large with respect to
observed values.  A natural sequel of the N14 work should address
the questions of whether this gas can possibly lead to SF, when SF
takes place, and what is the fate of the cold discs, whether they are
consumed or they are continuously replenished by cooling hot gas.  In
this paper we add SF to the simulations of N14, and we explore the ISM
evolution including the removal of cold gas, and the injection of
mass, momentum and energy appropriate for the newly (and continuously)
forming stellar population. In this way we aim at establishing whether
i) the N14 results for the general trends of the hot gas properties
with galaxy shape and stellar kinematics still hold; ii) the formation
of stars can reduce the amount of cold gas in the simulations, thus
bringing it more in agreement with observed values; iii) a significant
channel for SF, previously neglected, should be taken into
consideration for rotating systems, and whether this can account for
the low level SF activity currently seen to be ongoing.

This paper is organized as follows. In Section 2 we describe the main
ingredients of the simulations, such as the galaxy models and the
input physics.  In Section 3 we present and discuss the results of the
simulations. In Section 4 we summarize the conclusions.

\section{The simulations}

N14 performed a large set of 2D hydrodynamical simulations with the
ZEUS MP2 code to fully explore the large parameter space of realistic
(axisymmetric) galaxy models, characterized by different stellar mass,
intrinsic flattening, distribution of dark matter, and internal
kinematics. The galaxy flattening was either fully supported by ordered rotation,
originating the set of models that are isotropic rotators, or by
tangential anisotropy, originating the set of fully velocity
dispersion supported models. These two extreme configurations were built adopting the Satoh 
decomposition, respectively with Satoh parameter $k=1$ and $k=0$.
The galaxy models were tailored to
reproduce the observed properties and scaling laws of early type
galaxies \citep[see also][]{posacki.etal2013}.  In this work we perform
hydrodynamical simulations for a representative subset of rotating
models already investigated by N14 (Sect. 2.1), including SF in the
simulations (as described in Sect. 2.2 below), but keeping the code
(numerical set-up, grid properties) in all equal to that used by N14 (see N14 for details). 
Also, a less extreme value of $k=0.1$ is explored.
\referee{A logarithmically spaced numerical mesh $(R,z)$ of
  960$\times$480 gridpoints is employed, 
with a resolution of 90 pc in the first 10 kpc from the centre.}

%%%%%%%%%%%%%%%%%%%%%%%%%%%%%%%%%%%%%%%%%%%%%%%%%%%%%%%%%%%%%%%%%%%%%%
\renewcommand\arraystretch{1.4}
\begin{table*}
\caption{Main properties of the galaxy models.}
\begin{tabular}{ccccccccc}
\toprule
Name                             &$L_B$     &$R_{\rm e}$  &$M_*$                &$M_\mathrm{h}$             &   $\sigma_{\rm e8}$    &$f_{\rm DM}$ & $c$      \\ 
                             &$(10^{11}$L$_{\sun})$& (kpc) &$(10^{11}$M$_{\sun})$&$(10^{11}$M$_{\sun})$ &(km s$^{-1}$)&       &            \\
(1)                             & (2)           & (3)           & (4)    &   (5)                  &   (6)          & (7)    & (8)    \\
%E0$^{200}$                           &0.27           &4.09  &1.25             &25              & 200    &200     & 0.61   &   0.57 &   37  \\
\midrule                                                                   
EO4$^{200}_{\rm{IS}}$                &0.26           &4.09  &1.25               &25              & 166       & 0.63   &   37     \\
EO7$^{200}_{\rm{IS}}$                &0.26           &4.09  &1.25               &25              & 124     & 0.66   &  37     \\
FO4$^{200}_{\rm{IS}}$                &0.26           &4.09  &1.25               &25              & 178      & 0.59   & 37     \\
FO7$^{200}_{\rm{IS}}$                &0.26           &4.09  &1.25               &25              & 150      & 0.57   &  37     \\
\midrule                                                                   
%E0$^{250}$                           &0.65           &7.04  &3.35             &67              & 250    &250     & 0.59   &   0.55 &   28 \\
%\midrule                                                                   
EO4$^{250}_{\rm{IS}}$                &0.62           &7.04  &3.35               &67              & 207    & 0.62   &     28     \\
%EO4$^{250}_{\rm{VD}}$                &0.65           &7.04  &3.35               &67              & 223    &224     & 0.62   &   0.57 &   28     \\
EO7$^{250}_{\rm{IS}}$                &0.62           &7.04  &3.35               &67              & 154      & 0.66   &    28     \\
%EO7$^{250}_{\rm{VD}}$                &0.65           &7.04  &3.35               &67              & 184    &185     & 0.66   &   0.61 &   28     \\
%FO4$^{250}_{\rm{IS}}$                &0.65           &7.04  &3.35               &67              & 223    &224     & 0.57   &   0.53 &   28     \\
%FO4$^{250}_{\rm{VD}}$                &0.65           &7.04  &3.35               &67              & 240    &241     & 0.57   &   0.53 &   28     \\
%FO7$^{250}_{\rm{IS}}$                &0.65           &7.04  &3.35               &67              & 189    &190     & 0.56   &   0.51 &   28     \\
%FO7$^{250}_{\rm{VD}}$                &0.65           &7.04  &3.35               &67              & 223    &224     & 0.56   &   0.51 &   28     \\
%\midrule                                                                   
%\midrule
%E0$^{300}$                           &1.38           &11.79 &7.80             &160             & 300    &300     & 0.62   &   0.57 &   22   \\
%\midrule                                                                
\midrule                                                                   
EO4$^{300}_{\rm{IS}}$                &1.32           &11.79 &7.80               &160             & 248   & 0.64   &  22     \\
EO7$^{300}_{\rm{IS}}$                &1.32           &11.79 &7.80               &160             & 185   & 0.68   &  22     \\
FO4$^{300}_{\rm{IS}}$                &1.32           &11.79 &7.80               &160             & 266    & 0.60   &  22     \\
FO7$^{300}_{\rm{IS}}$                &1.32           &11.79 &7.80               &160             & 224    & 0.59   &    22     \\
\bottomrule
\end{tabular}
\flushleft
\textit{Notes}. $(1)$ Model name: ``EO'' or ``FO'' indicate the procedure (EO-building or FO-building) 
applied to the spherical (E0) model to obtain the shape indicated by the number (E4 or E7 shapes);
the superscript indicates $\sigma_{\rm e8}$ of the corresponding E0 model; 
the subscript ``IS'' indicates the kinematical configuration of the isotropic rotator
(for example, FO4$^{200}_{\rm{IS}}$ is an E4 isotropic rotator, FO-built from a spherical model with $\sigma_{\rm e8}=200$ km s$^{-1}$).
$(2)$ Luminosities in the $B$ band.
$(3)$ Effective radius (for a FO view for FO-built models, and an EO view for
EO-built models). 
$(4)$ Total stellar mass of the original (old) stellar population. The first four models are LM models, the next two are intermediate mass models, 
the last four are HM models.
$(5)$ Total dark matter mass.
$(6)$ Luminosity-weighted average of the  stellar velocity dispersion
within a circular aperture of radius $R_{\rm e/8}$;
for non-spherical models, $\sigma_{\rm e8}$ is the edge-on viewed value.
$(7)$ Dark matter fraction enclosed within a sphere of radius $R_{\rm e}$.
$(8)$ Concentration parameter of the dark matter profile. \\
Note: the properties listed above, for the EO4$^{250}$ and EO7$^{250}$ models with $k=0.1$, are not reported, since they are equal 
to those of the EO4$^{250}_{\rm {IS}}$
and EO7$^{250}_{\rm {IS}}$ ones, except for $\sigma_{e8}$, which is respectively 223 km s$^{-1}$ for the EO4$^{250}_{k\rm{=0.1}}$ model,
and 184 km s$^{-1}$ for the EO7$^{250}_{k\rm{=0.1}}$ model.
\label{tab1}
\end{table*}
\renewcommand\arraystretch{1.}

\subsection{The galaxy models}

N14 built axisymmetric two-component galaxy models where the stellar
component has two different intrinsic flattening, corresponding to the
E4 and E7 shapes, while the dark matter halo is spherical. The
luminous matter is described by the deprojection \citep{mellier.etal1987} of the
\citet{devaucouleurs.1948} law, generalized for ellipsoidal
axisymmetric distributions; the dark matter profile is the
\citep{navarro.etal1997} one, with the dark mass $M_\mathrm{h}$ amounting at $\simeq 20$
times the total
stellar mass $M_*$.  For any fixed galaxy mass and shape (E4 or E7), N14 built two models:
the first one, called ``FO-built'', when seen
face-on has the same $R_{\rm e}$ of the spherical E0
counterpart, thus its stellar mass distribution becomes more and more
concentrated than in the E0 model, as it gets flatter; the second one, called ``EO-built'',
when seen edge-on has the same circularized $R_{\rm e}$ of the E0
counterpart, which makes its stellar mass distribution to
expand with increasing flattening.

In this work we re-simulate a few flat (E4 and E7) rotating models
(isotropic rotators, with $k=1$) of N14 including SF in them. In order to
explore the effects of SF at the high and low ends of the galaxy mass
range explored by N14, we choose four flat models with luminosity-weighted stellar
velocity dispersion within $R_{\rm e/8}$ of $\sigma_{\rm e8} = 300$ km
s$^{-1}$ for the parent spherical model (and we call these high-mass models, ``HM'' models), 
and four with $\sigma_{\rm e8} = 200$ km s$^{-1}$ for the E0 counterpart (low-mass models, ``LM'' models).  In N14, the first
set was found to host inflows, with the creation of a massive,
centrifugally supported cold gaseous disc, while the second set was found in
a global wind, or close to the transition to it. 
\referee{In a global wind the gas has very low density and positive velocity (directed outwards) through most of the galaxy
(e.g., \citealt{mathews.baker1971})\footnote{ In fact the phase of the gas flow  (that can range from a wind to an inflow) is basically determined by the
different relative importance of SN heating in galaxies of different mass, as already 
thoroughly  discussed in \citealt{ciottietal1991}.}.}
In addition, we
re-simulated two intermediate-mass models (again E4 and E7), with
$\sigma_{\rm e8} = 250$ km s$^{-1}$ for the E0 counterpart.  For these two models, 
we also built a moderately rotating stellar kinematical configuration ($k=0.1$), not explored by N14.
The main structural properties of the ten re-simulated
models, identical to those presented in N14, are listed in Tab. 1; 
the two new models with $k=0.1$ differ
from the intermediate-mass models only in the $\sigma_{\rm e8}$ value, which is given in the notes to the table.

\subsection{Star formation in the code}

We employed two different schemes for SF, a passive (pSF) one and an
active (aSF) one.  In the pSF, only cold gas removal from the
numerical mesh is allowed; in the aSF, we also consider the injection
of mass, momentum and energy from the newly forming stellar
population, for simplicity limiting in this work to the evolution of stars more massive than
8~M$_{\sun}$ (ending with SNII explosions); note however that for reasonable stellar initial mass
functions (IMF) these massive stars are major contributors to the total mass return rate.   
We describe below the
main inputs and sinks of mass and energy for the gas flow, due to both the original
stellar population (of total mass $M_*$) and the newly forming stars, and the  
corresponding equations of hydrodynamics solved by the code.

\subsubsection{The  mass injection and sink terms}\label{mass1}

The mass inputs are stellar winds and SNIa's ejecta produced during 
the passive evolution of the original stellar population of the
galaxy (at a rate per unit volume respectively of $\dot\rho_*$ and
$\dot\rho_{\rm Ia}$, for which we adopt the standard recipes coming from
the stellar evolution theory; e.g., N14), and the Type II
supernovae ejecta produced by the newly born stellar population (at a
rate of $\dot\rho_{\rm II}$, calculated as detailed below). 

The mass sink is due to SF subtracting gas from the grid, at an
adopted rate per unit volume of:
\begin{equation}
\dot\rho_{\rm SF} = \dfrac{\eta_{\rm SF} \rho}{t_{\rm SF}},\qquad 
t_{\rm SF}=\mathrm{max}(t_\mathrm{cool},t_\mathrm{dyn}), \label{eq:sf}
\end{equation}
where $\rho$ is the gas density, and
$\eta_{\rm SF}$ is the SF efficiency, for which we adopt two values of $\eta_{\rm SF}=0.01$
and $0.1$. The rate $\dot\rho_{\rm SF}$ depends on the maximum between the cooling
timescale $t_\mathrm{cool}\equiv E/\cal{L}$ (where $E$ is the ISM internal
energy density, and $\cal{L}$ is the ISM bolometric luminosity per unit volume),
and the dynamical timescale $t_\mathrm{dyn} \equiv \sqrt{ 3\upi/32 G\rho} $
\citep[see also][]{ciotti.ostriker2007}. 

When SF takes place in a computational cell, the ISM is removed and an
equal mass in stars $\Delta M_*$ appears at the same place.  These new
stars are assumed to form with a Salpeter IMF, thus for a given
$\Delta M_*$, the number of stars having a mass greater than
8~M$_{\sun}$, that will explode as SNII, is $ N_{\rm II}\simeq 7\times
10^{-3} \Delta M_*($M$_{\sun})$.  These new stars  in turn inject mass into the
ISM.  By integration of the mass difference between the mass of the
progenitor and that of the remnant, for the Salpeter IMF, one finds that the mass
injected by SNII's is $\simeq 0.2 \Delta M_*$.  The final SNII mass
source term $\dot \rho_{\rm II}$ comes from considering that a given SF
episode generates SNII's that inject mass at a rate exponentially
declining on a timescale $\tau_{\rm II} = 2\times 10^7$ yr, and that at a
certain time during the evolution of a SF episode, another episode may
take place, forming younger SNII's that in turn eject mass into the
ISM. The whole process can be formalized as:
\begin{equation}
\dfrac{\diff \dot\rho_{\rm II}}{\diff t} = -\dfrac{\dot\rho_{\rm II}}{\tau_{\rm II}
} +
\dfrac{0.2 \dot\rho_{\rm SF}}{\tau_{\rm II}}, \label{rhoII}
\end{equation}
\citep{ciotti.ostriker2007}. The equation above is very useful in numerical works, as it allows to compute the mass return 
of the new stars formed at each timestep
without storing the whole SF history at each gridpoint, but only the current value and the value of the mass return at the previous timestep.
For more complex recipes, built on the scheme above, see 
\citet{calura.etal2014}.

\subsubsection{The energy injection and sink terms}

Energy is injected into the ISM by the thermalization of the kinetic energy of SNIa and SNII explosions, and
by the thermalization of random and streaming stellar motions for the stellar winds. 
The rate of SNIa's explosions is the same
entering $\dot\rho_\mathrm{Ia}$ defined in Sect.~\ref{mass1}.
The energy input rate per unit volume from the original (old) stellar population is then:
\begin{equation}
{\dot E}_{\mathrm{old}} = {\dot E}_{\mathrm{Ia}} + \dfrac{\dot\rho_{\rm
Ia}+\dot\rho_*}{2} \left[
\norma{\vvphi {\ephi} -{\boldsymbol{u}}}^2 + \Tr ({\boldsymbol{\sigma}}^2)
\right],\label{eold}
\end{equation}
where ${\dot E}_{\mathrm{Ia}} = \dot\rho_\mathrm{Ia}\vartheta_\mathrm{SNIa} E_\mathrm{SN}/{\rm 1.4~M_{\sun}} $,
$\vartheta_\mathrm{SNIa}$ is the thermalization efficiency \referee{\citep[for which we adopt the value of 0.85, see ][]{thornton.etal1998, tang.wang2005}}, $E_\mathrm{SN}=10^{51}$ erg,
and 1.4~M$_{\sun}$ is the mass that is ejected by one SNIa event; $\vvphi$ is the stellar streaming velocity field, 
${\boldsymbol{u}}$ is the velocity of the ambient gas, 
$\boldsymbol{\sigma}^2$ is the stellar velocity dispersion tensor of the stars.

The energy input rate ${\dot E}_\mathrm{new}$ from the newly born stellar population is derived as follows.
A mass of newly formed stars $\Delta M_*$ injects energy through SNII with an efficiency
\begin{equation}
 \varepsilon_\mathrm{II} = \dfrac{N_\mathrm{II}  E_\mathrm{SN}}{\Delta M_* c^2}
\simeq 3.9\times 10^{-6}.
\end{equation}
Consistently with eq.~\ref{rhoII}, the SNII energy injection rate ${\dot E}_{\rm II}$ 
due to the thermalization of the ejecta is given by:
\begin{equation}
\dfrac{d {\dot E}_{\rm II}}{dt} = -\dfrac{\dot E_{\rm II}}{\tau_{\rm II} } +
\dfrac{\varepsilon_\mathrm{II} c^2  \dot\rho_{\rm SF}
\vartheta_\mathrm{SNII}}{\tau_{\rm II}}, \label{eII}
\end{equation}
where we take the fiducial value of $\vartheta_\mathrm{SNII}=\vartheta_\mathrm{SNIa}/5=0.17$ 
to account for the lower thermalization efficiency of SNII exploding in a cold and dense medium.
We assume that the
new stars inherit the kinematical configuration of the original stellar
component in the place where they are born (i.e. the velocity dispersion and rotation of the new stars are the same as those of 
the original stellar distribution in the same place).
\referee{This allows us to treat the heating terms described in eq. (7) below by using the same properties of the old stars. Note however
that this choice is not unreasonable, because the bulk of star formation takes place in the cold and rotationally supported disc,
whose rotational properties are very similar to those of the stars.} 
Thus, the total heating rate per unit volume due to Type II SNe is
\begin{equation}
\dot E_\mathrm{new} = {\dot E}_{\rm II} 
+ \dfrac{\dot\rho_{\rm II}}{2} \left[ \norma{\vvphi \ephi -{\boldsymbol{u}}}^2 +
\Tr ({\boldsymbol{\sigma}}^2)
\right].
\end{equation}
The total energy injection due to the old and new stellar populations is then
$\dot E = \dot E_\mathrm{old} + \dot E_\mathrm{new}$, that is:
\begin{equation}
\dot E =  {\dot E}_{\mathrm{Ia}}  +
 {\dot E}_{\mathrm{II}}  + \dfrac{\dot\rho_{\mathrm{Ia}}+\dot\rho_* + \dot\rho
_{\mathrm{II}}}{2} \left[
\norma{\vvphi {\ephi} -{\boldsymbol{u}}}^2 + \Tr ({\boldsymbol{\sigma}}^2)
\right] . \label{edot}
\end{equation}
Finally, \referee{an energy and momentum sink associated with SF are present; 
they are respectively written as}:
\begin{equation}
{\dot E}_{\rm SF} = \dfrac{\eta_{\rm SF} E}{t_{\rm SF} }, \quad\quad\quad
{\dot{{\boldsymbol{m}}}_\mathrm{SF}} = \dfrac{\eta_{\rm SF}
{\boldsymbol{m}}}{t_{\rm SF} }, \label{eq:star_form_edot}
\end{equation}
\referee{where $E$ and $\boldsymbol{m}$ are the internal energy and momentum density of
the ISM.}
%We assume that the sink velocity of the starforming gas is the local ISM
% velocity $ {\boldsymbol{u}}$.

\subsubsection{The hydrodynamical equations}

The hydrodynamical equations are the same as in N14, with the addition 
of all the source and sink terms related with the SF process described above:
\begin{gather*}
%\begin{equation}
 \dtpartial{\rho} + \diver (\rho {\boldsymbol{u}}) = \dot\rho_\mathrm{Ia} +
\dot\rho_* +
\dot\rho_{\rm II} - \dot\rho_{\rm SF},\label{eq:uff} \\[2ex]
\begin{split}
\rho \dtpartial{\boldsymbol{u}} + \rho
\convective{\boldsymbol{u}}{\boldsymbol{u}} = &-\grad p
-\rho\grad\Phi_\mathrm{tot} +\\
&+ (\dot\rho_\mathrm{Ia} + \dot\rho_* +\dot\rho_{\rm II} ) (\vvphi \ephi -
{\boldsymbol{u}}) ,
\end{split}\\[2ex]
\dtpartial{E} + \diver (E{\boldsymbol{u}}) = -p \diver {\boldsymbol{u}} -
{\mathcal{L}}
+  \dot E -\dot E_\mathrm{SF}, \label{eq:hydroeqs}
%\end{equation}
\end{gather*}
where $\rho$, ${\boldsymbol{u}}$, $E$, $p$, $\Phi_{\rm tot}$, and
${\cal L}$ are respectively the ISM mass density, velocity, internal energy
density, pressure, total gravitational potential, and bolometric luminosity
per unit volume. The gas is assumed to be an ideal
monoatomic fully ionized plasma, so that $p = (\gamma -1)E$, where $\gamma =
5/3$ is the adiabatic index. The chemical composition is fixed to solar
($\mu\simeq 0.62$), and the gas self-gravity is neglected.

In the pSF scheme, the source terms $\dot\rho_{\rm II}$ and ${\dot
E}_{\mathrm{II}}$  
are zero, thus they either do not enter the hydrodynamical 
equations above and they do not contribute to $\dot E$. In the more realistic
aSF, $\dot\rho_{\rm II}$ and ${\dot E}_{\mathrm{II}}$ are non-zero. 
Most galaxy models are simulated with pSF and aSF; for each of the two schemes, we adopted
two values of the SF efficiency in eqs.~\ref{eq:sf} and~\ref{eq:star_form_edot},
$\eta_{\rm SF}=10^{-1}$ and $\eta_{\rm SF}=10^{-2}$ (see Sect. 3 below).

\begin{figure*}
\centering
\includegraphics[width=0.95\linewidth,keepaspectratio]{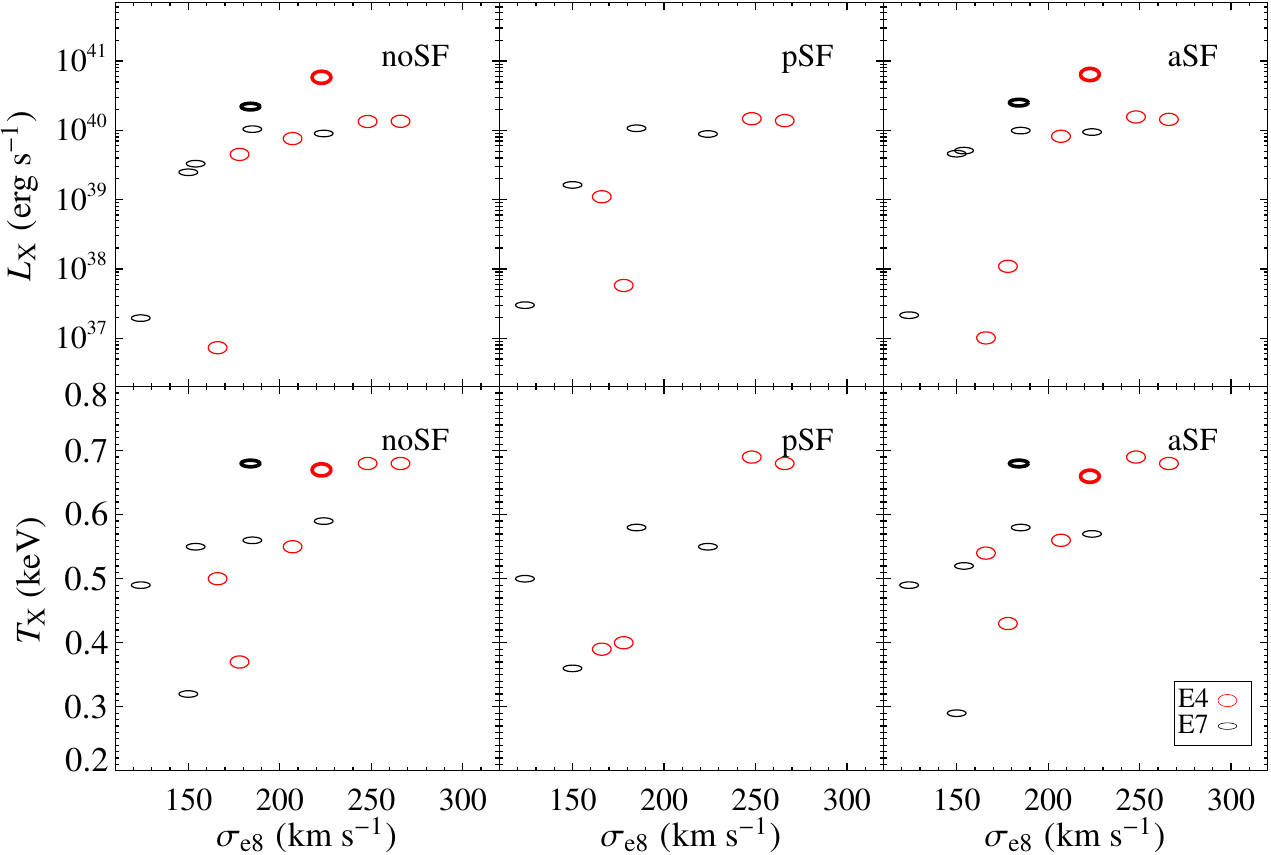}
\caption{Top panels: ISM X-ray luminosity $\Lx$ in the 0.3--8 keV band at 13 Gyr
as a function of $\sigma_{\rm e8}$, for the selection of ten models from N14  re-simulated here 
(Tab.~\ref{tab1}), plus two models with $\sigma_{\rm e8}=250$ km s$^{-1}$ for the E0 counterpart, 
and $k=0.1$ (shown with thicker ellipses).
The three panels (from left to right) refer to the same models without SF (from N14), with passive SF, and with active SF
(with $\eta_{\rm SF}=0.01$). 
Bottom panels: the same for the X-ray emission weighted temperature $\Tx$ in the 0.3--8 keV band at 13 Gyr.
All $L_X$ and $\Tx$ values are given in Tabs. \ref{tab2}-\ref{tab5}. See
Sect.~\ref{comp} for more details.}
\label{Lx}
\end{figure*}
\begin{figure}
\centering
\includegraphics[width=\linewidth,keepaspectratio]{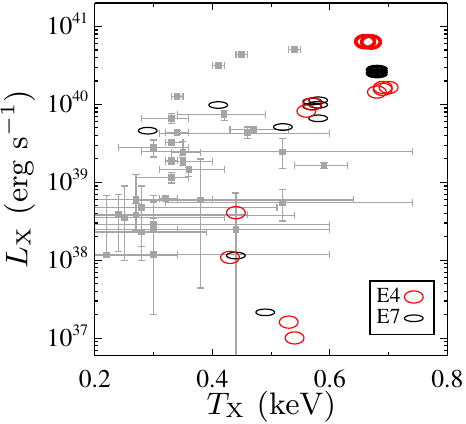}
\caption{$\Lx$ as a function of $\Tx$ for aSF models listed in Tables~\ref{tab1}-\ref{tab5} (open ellipses, as in Fig.~\ref{Lx}), compared with \textit{Chandra} data for the hot gas of ETGs taken from \citet{boroson.etal2011} and \citet{kim.fabbiano2015} (grey squares with error bars). From the latter two samples
we selected only ETGs of morphological type later than E3, for a proper comparison with our models.}
\label{fig2_new}
\end{figure}

We tested that the code conserves the total mass and energy; the total
mass at any time $t$ is $M_\mathrm{gas} +M_\mathrm{esc} = M_\mathrm{inj}
+M_\mathrm{inj}^{\rm II}-M_*^{\rm new}$, where $M_\mathrm{gas}$ is the gas mass within the
simulation box, $M_\mathrm{esc}$ is the cumulative escaped mass out of the
simulation box, until that time; $M_\mathrm{inj}$, $M_\mathrm{inj}^{\rm II}$, and
$M_*^{\rm new}$ are the cumulative masses respectively injected by the
passively evolving stellar population, injected by SNII events due to
the newly born stellar population, and of the new stars.  The values
of these masses at the end of the simulations are given in Tabs.
\ref{tab2}-\ref{tab5}.
Finally, the radiative cooling is implemented by adopting a modified
version of the cooling law of \citet{sazonov.etal2005},
with
a lower limit for the ISM temperature of $T = 10^4$ K.
The total X-ray emission in the 0.3–8 keV
band ($\Lx$), and the emission weighted temperature in the same band ($\Tx$), 
are calculated  as volume integrals over the whole computational grid, 
using as weight the emissivity in the 
0.3–8 keV band of a hot, collisionally ionized plasma (see N14 for more details).

\section{Results}

Each simulation starts with the galaxy empty of gas, when the age of
the original stellar population is 2 Gyr, and the ISM evolution is
followed for the next 11 Gyr.  Tables \ref{tab2}-\ref{tab5} list for all models
the main
quantities of interest at the end of the simulations; the
tables also list the same quantities for the corresponding models
without SF taken from N14 (hereafter ``noSF'' models).  First
(Sects.~\ref{comp}) we present the main results focussing on the two
sets of re-simulated models with largely differing values for the galaxy mass
(i.e., the two sets with E0 counterpart of $\sigma_{\rm e8} = 200$ and
300 km s$^{-1}$, the LM and HM models). In Sect.~\ref{interm}
we concentrate on the intermediate mass models (run only for the aSF), where this time
$k$ is equal to 1 or lower ($k=0.1$). In Sects.~\ref{mass}
and~\ref{hist} we
discuss the consumption of cold gas mass and the formation of new stars.
Finally, we investigate the relationship
between the adopted recipe for SF (eq.~\ref{eq:sf}) and the
Kennicutt-Schmidt \referee{relation} (Sect.~\ref{KS}).

\begin{figure*}
\includegraphics[angle=90,width=0.83\linewidth,keepaspectratio]{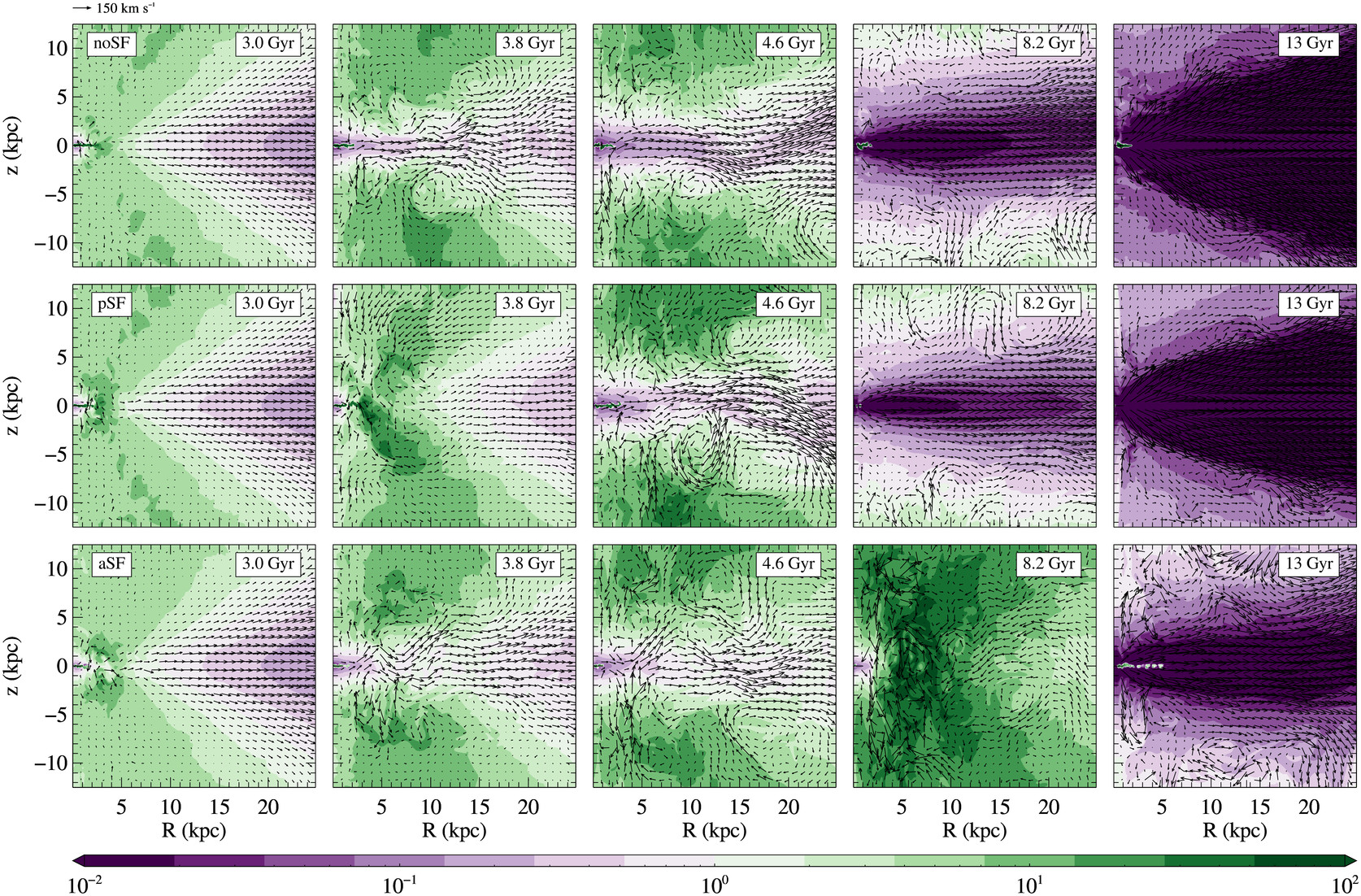}
\caption{Meridional sections of the heating over cooling time ratio for the LM EO7$^{200}_\mathrm{IS}$ model, 
at a selection of representative times (indicated as galaxy ages in the top right of each panel). 
The heating time is $t_\mathrm{heat}=E/\dot E$ (where $\dot E$ is the source term given in eq.~\ref{edot}); the cooling time is defined in Sect. 2.2.1.
From left to right, each column refers to the noSF, pSF and aSF (with $\eta_{\rm SF}= 0.1$) models, respectively. 
Arrows describe the meridional velocity field; their lenght
is proportional to the modulus of the gas velocity in the ($z,R)$ plane, according to a scale
shown in the bottom left corner. See Sect.~\ref{comp} for more details.}
\label{EO7}
\end{figure*}

\begin{figure*}
\includegraphics[width=0.85\linewidth,keepaspectratio]
{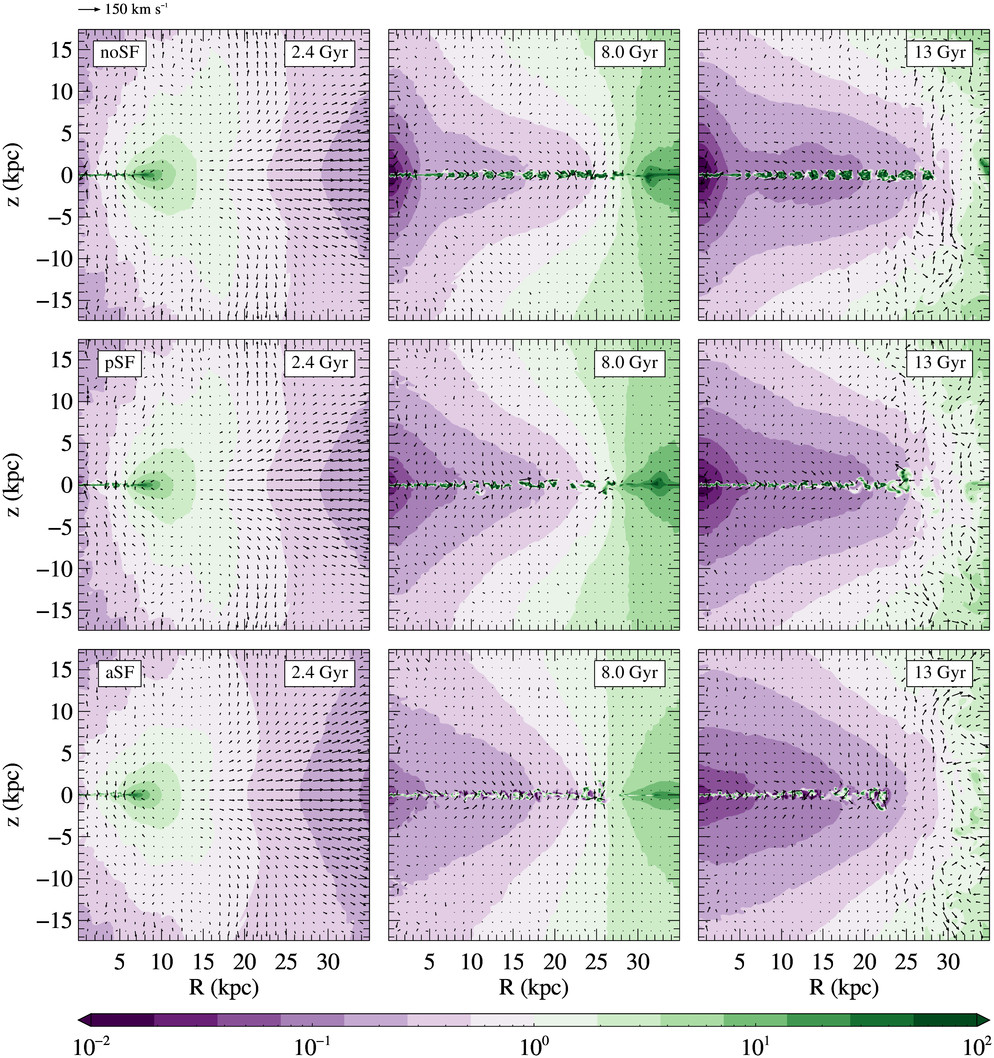}
\caption{Meridional sections of the heating over cooling time ratio for the HM
EO7$^{300}_\mathrm{IS}$ model. From top to bottom, each row refers to the noSF, pSF and aSF
(with $\eta_\mathrm{SF} = 0.1$) models,
respectively. Arrows describe the velocity field, as in Fig.~\ref{EO7}.
Note the large cold disc, that does not completely disappear with SF. See Sect.~\ref{comp} for more details.}
\label{s300}
\end{figure*}

\subsection{Comparison between noSF and SF models}\label{comp}
 
In a short summary, for noSF models, N14 found that rotation in LM
galaxies favours the establishment of global winds, with the
consequent reduction of $\Lx$; in medium-to-high mass galaxies the
conservation of angular momentum lowers the hot gas density in the
central galactic region, leading again to a reduction of $\Lx$, and
also of $\Tx$ (because the external and colder regions weight more in
the computation of $\Tx$).  In LM galaxies, instead, $\Tx$ can become
higher if rotation triggers a wind, due to the decrease of the ISM
density, and the additional heating due to the high meridional
velocities of the escaping material (see eq.~\ref{edot}). 

\begin{figure*}
%\subfloat{\includegraphics[width=0.86\linewidth,keepaspectratio]{LxTx_NFWi_200_IS_E7_EO}}\\
\subfloat{\includegraphics[width=0.8\linewidth,height=0.35\linewidth]{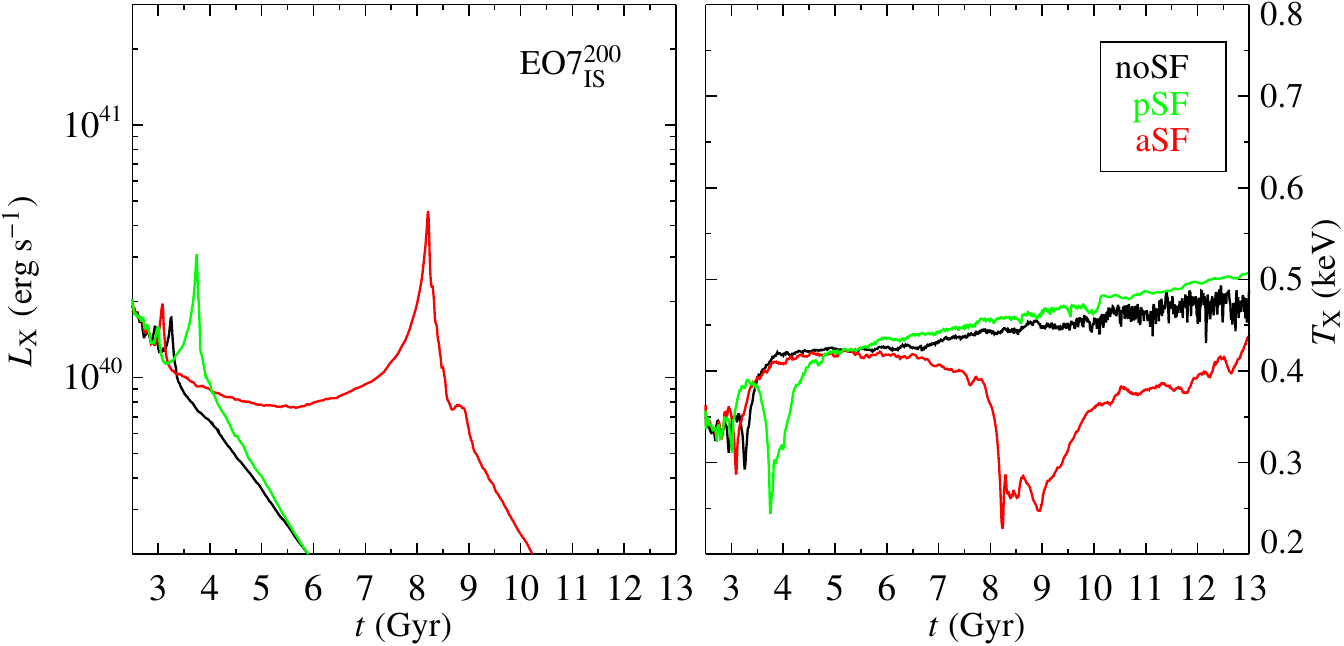}}\\
% 
%\subfloat{\includegraphics[width=0.86\linewidth,keepaspectratio]{LxTx_NFWi_200_IS_E7_FO}}\\
\subfloat{\includegraphics[width=0.8\linewidth,height=0.35\linewidth]{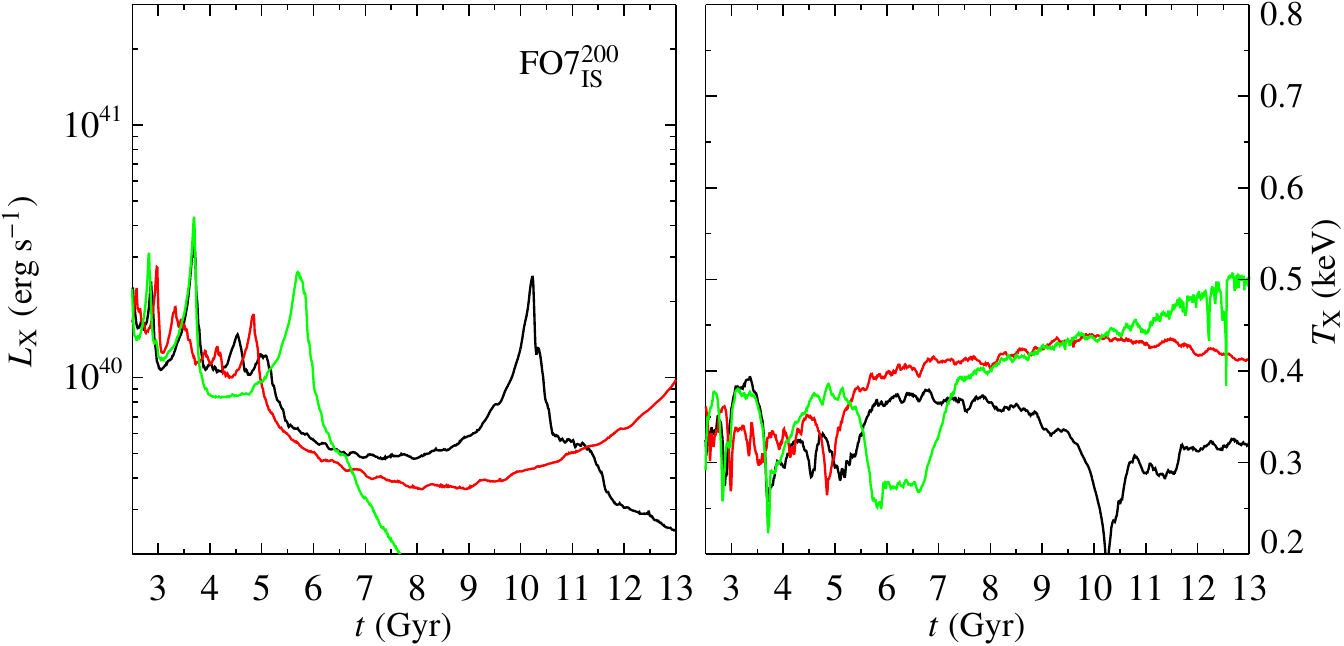}}\\
% 
%\subfloat{\includegraphics[width=0.86\linewidth,keepaspectratio]{LxTx_NFWi_300_IS_E7_EO}}
\subfloat{\includegraphics[width=0.8\linewidth,height=0.35\linewidth]{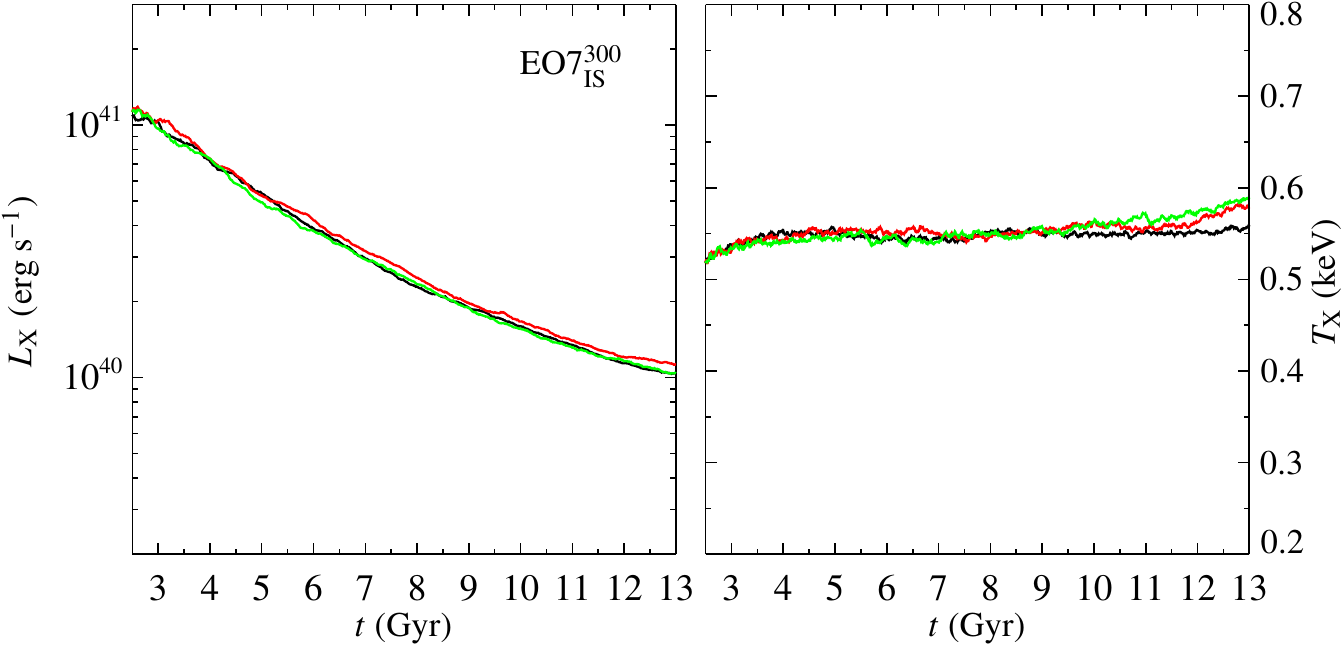}}
\caption{Time evolution of the X-ray luminosity $\Lx$ and X-ray emission
weighted temperature $\Tx$ for the EO7$^{200}_\mathrm{IS}$ (top panels),
FO7$^{200}_\mathrm{IS}$ (middle panels), EO7$^{300}_\mathrm{IS}$ (bottom panels)
models. The black, green and red lines refer to the noSF model, and to the passive and active star formation schemes, respectively, with $\eta_{\rm SF} =
0.1$. The FO7$^{300}_\mathrm{IS}$ model has an evolution  almost
identical to that of EO7$^{300}_\mathrm{IS}$ shown here.  See Sect.~\ref{comp} for more details.}
\label{LTt}
\end{figure*}

The first important result here is that when SF is added (both in the pSF and aSF modalities), the hot ISM
evolution remains substantially similar to that found by N14 without
SF, for models of the same mass.  
As a consequence of the insensitivity of the general behaviour to the addition 
of SF, the values of $\Lx$ and $\Tx$ for the models with SF, and
their trends with the main galactic structural properties, 
are on average very similar to those found by N14. Therefore, the conclusions of N14 about
the importance of shape and rotation in determining $\Lx$ and $\Tx$,
also as a function of galaxy mass, are confirmed.
Figure~\ref{Lx} demonstrates this
result by plotting the final $\Lx$ and $\Tx$ values for the same rotating galaxy models evolved
with noSF and with SF (with $\eta_{\rm SF}=0.01$): the
distribution of the points in the noSF and SF panels
is very similar (and the analogous figure with the $\eta_{\rm SF}=0.1$ models produces the same conclusions). 
In particular, $\Lx$ of HM models is only marginally sensitive to the presence of SF, 
while in LM models SF can introduce variations of $\Lx$; however, these keep within 
the (large) range of $\Lx$ values already found in N14 without SF. Thus, SF adds another cause of spread for $\Lx$,
at lower galaxy masses.
\referee{Figure~\ref{fig2_new} shows the resulting $\Lx$ vs. $\Tx$ for the aSF models, compared with the same quantities derived
recently from \textit{Chandra} data for the hot gas of ETGs (\citealt{boroson.etal2011} and \citealt{kim.fabbiano2015}). 
From the latter two samples
we have selected only ETGs of morphological type later than E3, for a proper comparison with our models. 
Overall, the simulation results agree well with the observed X-ray properties of real ETGs (except perhaps for the most X-ray luminous models
that may be slightly hotter than observed).}

The reason for variations in $\Lx$ linked with SF in LM models is
given by anticipations or delays in the most conspicuous features of
their typical flow behaviour: \referee{N14 found (and the present simulations confirm) that the ISM in rotating models experiences 
periodic cooling episodes, where the gas injected by the stellar population accumulates until the radiative losses become catastrophic. 
In these episodes, the ISM quickly cools, emitting a large part of its internal energy as radiation; peaks in $\Lx$ and throats in $\Tx$ 
are then produced, as apparent in Fig. 5. The tenuous hot atmosphere left after a major cooling is then replenished by the new mass 
injection from the stellar population. In correspondence of the formation of these significant amounts of cold material (with short cooling 
times) star formation is enhanced, and peaks in pace with $\Lx$, as discussed in Sect. 3.4.}
On the contrary, very
little variations of the flow due to SF are seen in HM
galaxies. Figures~\ref{EO7}, \ref{s300}, and \ref{LTt} give particular
examples of the general similarity in the hydrodynamical evolution of
noSF and SF models, by plotting in parallel the ISM evolution for the
noSF, pSF and aSF cases; at the same time, the figures point out how
this similarity is very close for HM models (see Fig.~\ref{s300}, for
the HM model EO7$^{300}_{\rm IS}$), while some differences can be
present in LM ones (see Fig.~\ref{EO7}, for the LM model
EO7$^{200}_{\rm IS}$).  Notably, the hot ISM evolution in the HM
models is practically identical with and without SF (see also
Fig.~\ref{LTt}): a massive cold disc forms early, and afterwards, even
when it is mostly consumed by SF (see Sect.~\ref{mass} below), around
the disc the hot ISM evolution proceeds almost unaltered by the
energy and mass injection from SNII's.  Less massive models, instead,
are sensitive to the inclusion of SF, similarly to what found in
previous studies where they showed to be very sensitive to any change
in the galaxy properties (e.g., the mass distribution, the stellar
population inputs, the stellar kinematics; \citealt{ciottietal1991},
N14). As shown by Figs.~\ref{EO7} and \ref{LTt}, for the
EO7$^{200}_{\rm IS}$ galaxy, at around 4 Gyr the noSF and the pSF
models have already past their major cooling episode (see the peak in
$\Lx$ in Fig.~\ref{LTt}) followed by the onset of a wind, and the
equatorial outflow becomes stronger and stronger thereafter (see how
the purple regions become more and more extended in Fig.~\ref{EO7},
starting from 4.6 Gyr,
and the sharp and steady decline in $\Lx$ in Fig.~\ref{LTt}). The aSF
model, instead, has its major cooling catastrophe much later,  as
shown by the prominent dark green region in Fig.~\ref{EO7} at $\simeq
8$ Gyr, \referee{which emits a conspicuous amount of radiation in the X-ray band, thus prompting} the corresponding peak in $\Lx$ in Fig.~\ref{LTt}.
This delay in the major cooling episode, preceding the onset of a wind,
is due to the newly formed stars that inject mass, and most
importantly heat, in the ISM; thanks to this heat the galaxy can
remain hot gas rich until $\simeq 8$ Gyr, when finally the major
cooling takes place. pSF, where mass is not injected instead,
corresponds to an evolution very similar to that without SF.  

A complementary illustration of the different behaviour between the HM
and the LM galaxies is given by Fig.~\ref{LTt}, that shows the $\Lx$
and $\Tx$ evolution for the same models of Figs.~\ref{EO7}
and~\ref{s300}, plus another set of three models representative of the
LM class, the FO7$^{200}_{\rm IS}$ ones.  In LM models, a number of
cooling episodes (peaks in $\Lx$) typically take place, possibly terminated by a major
one that is followed by the onset of a wind and the clearing of the
ISM from the galaxy (as in N14).  The galaxy mass distribution in the FO7$_{\rm IS}^{\rm 200}$
models (middle panels of Fig.~\ref{LTt}), is more concentrated 
than in the EO7$^{200}_{\rm IS}$ ones, thus the last major peak in $\Lx$ is delayed
with respect to what happens for the EO7$^{200}_{\rm IS}$ models. Thus,
in the FO7$_{\rm IS}^{\rm 200}$ galaxy, aSF produces  a delay in the
major cooling episode that is even longer than for the EO7$_{\rm
  IS}^{\rm 200}$ aSF model, to the point that it does not take place
within the present epoch. pSF instead produces its anticipation with
respect to what shown in the noSF case (Fig.~\ref{LTt}),  due
to pSF subtracting gas and leaving a lower density region, radiating
less and more easy to push out of the galaxy.

A final, general result is also that the evolution of a model with SF is
more and more different from that of its corresponding noSF model, when
increasing $\eta_{\rm SF}$.

\begin{figure*}
\includegraphics[width=0.75\linewidth,keepaspectratio]{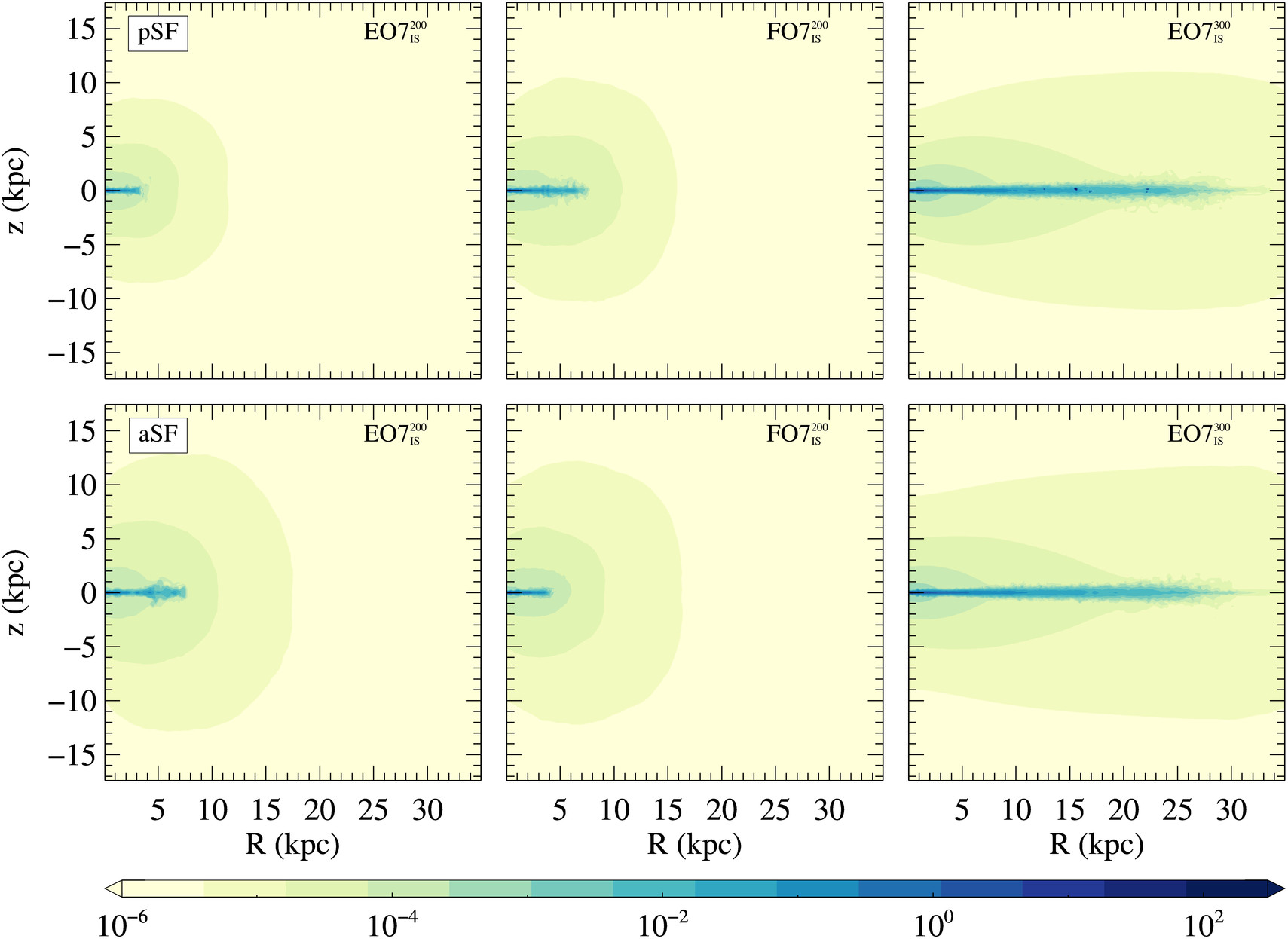}
%%\subfloat{\includegraphics[width=0.33\linewidth,keepaspectratio]{
% EO7-200ISP01newdenstar_000_time_1_2999e+10}}
% 
%%\subfloat{\includegraphics[width=0.33\linewidth,keepaspectratio]{
% FO7-200ISP01newdenstar_000_time_1_2999e+10}}
%
%%\subfloat{\includegraphics[width=0.33\linewidth,keepaspectratio]{
% EO7-300ISP01newdenstar_000_time_1_2999e+10}}\\
%  
%
%\subfloat{\includegraphics[width=0.33\linewidth,keepaspectratio]{
% EO7-200ISA01newdenstar_000_time_1_2999e+10}}
% 
%%\subfloat{\includegraphics[width=0.33\linewidth,keepaspectratio]{
% FO7-200ISA01newdenstar_000_time_1_2999e+10}}
% 
%%\subfloat{\includegraphics[width=0.33\linewidth,keepaspectratio]{
% EO7-300ISA01newdenstar_000_time_1_2999e+10}}\\
% 
\caption{Meridional sections of the density distribution of the
new stars formed up to 13 Gyr (in units of M$_{\sun}$ pc$^{-3}$), for the same models
of Figs.~\ref{LTt} and~\ref{Mt}:
EO7$^{200}_\mathrm{IS}$ (left panels), FO7$^{200}_\mathrm{IS}$ (middle panels),
and EO7$^{300}_\mathrm{IS}$ models (right panels). Top and bottom panels refer respectively
to pSF and aSF simulations, with $\eta_{\rm SF}= 0.1$. }
\label{newst}
\end{figure*}

For illustration purposes, Fig.~\ref{newst} shows a meridional section
of the density distribution of the newly formed stars at the end of
the simulations, for the models in Fig.~\ref{LTt}. One can notice the
more extended and massive disc in HM models, where it is also quite
independent of the pSF or aSF scheme. On the contrary, in LM models,
the disc is much less extended, and it can become larger (for the
EO7$^{200}_\mathrm{IS}$ model) or smaller (for the
FO7$^{300}_\mathrm{IS}$ model) when switching from the pSF to the aSF
scheme; this is another evidence of the high sensitivity of the flow
evolution to any change in the input parameters, at low galaxy masses.

\begin{figure*}
%\subfloat{\includegraphics[width=0.8\linewidth,height=0.35\linewidth]{
% LxTx_NFWi_250_IS_EO}}\\
%\subfloat{\includegraphics[width=0.8\linewidth,height=0.35\linewidth]{
% LxTx_NFWi_250_IS_EO_k01}}
\subfloat{\includegraphics[width=0.65\linewidth,height=0.55\linewidth]{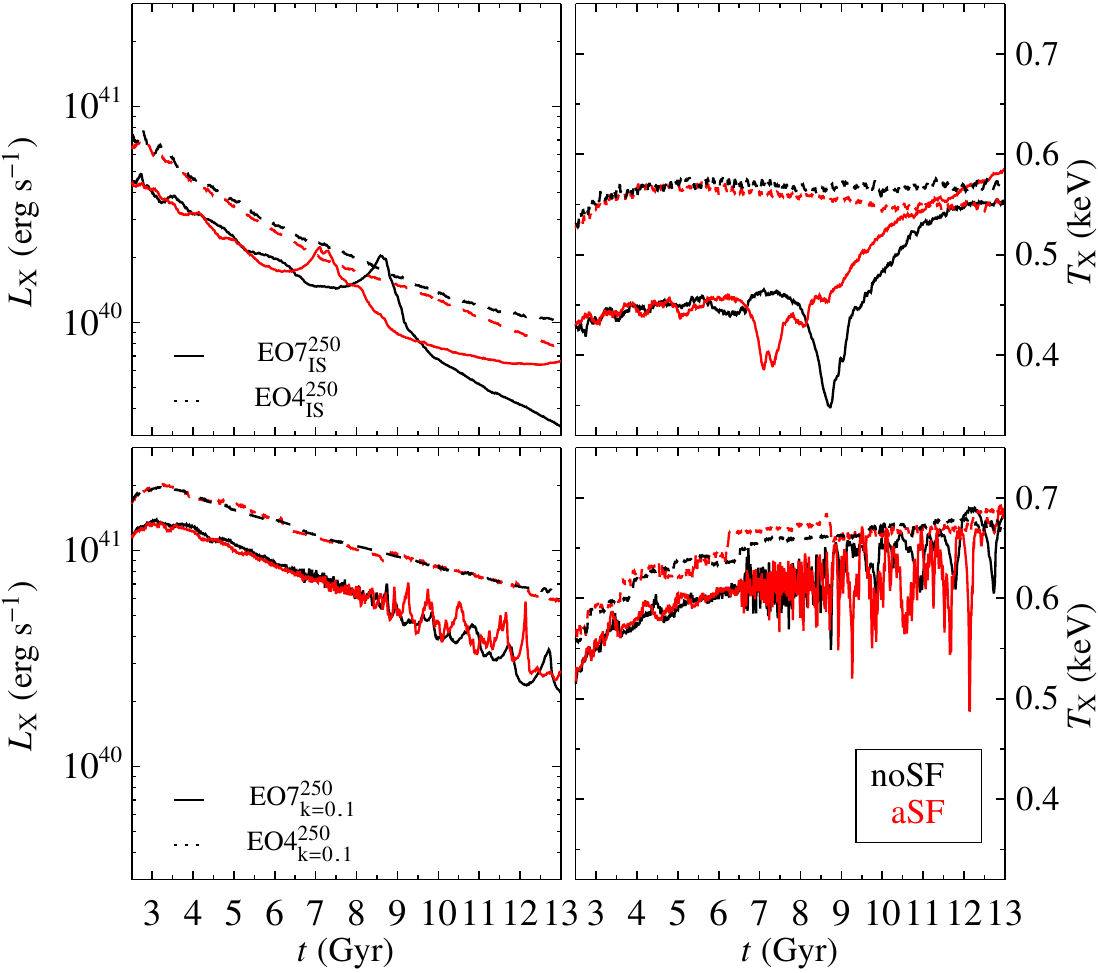}}
\caption{Time evolution of the X-ray luminosity $\Lx$ and X-ray emission
weighted temperature $\Tx$ for the EO4$^{250}_\mathrm{IS}$ and EO7$^{250}_\mathrm{IS}$ models
(top panels), and for the EO4$^{250}_{k=0.1}$ and EO7$^{250}_{k=0.1}$ models (bottom panels). 
For the aSF models, $\eta_{\rm SF} =0.1$. See Sect.~\ref{interm} for more details.}
\label{s250LT}
\end{figure*}

\begin{figure*}
%\subfloat{\includegraphics[width=0.67916666\linewidth,keepaspectratio]{paper3_mass_NFWi_200_IS_E7_EO}}\\
\subfloat{\includegraphics[width=0.8\linewidth,height=0.35\linewidth]{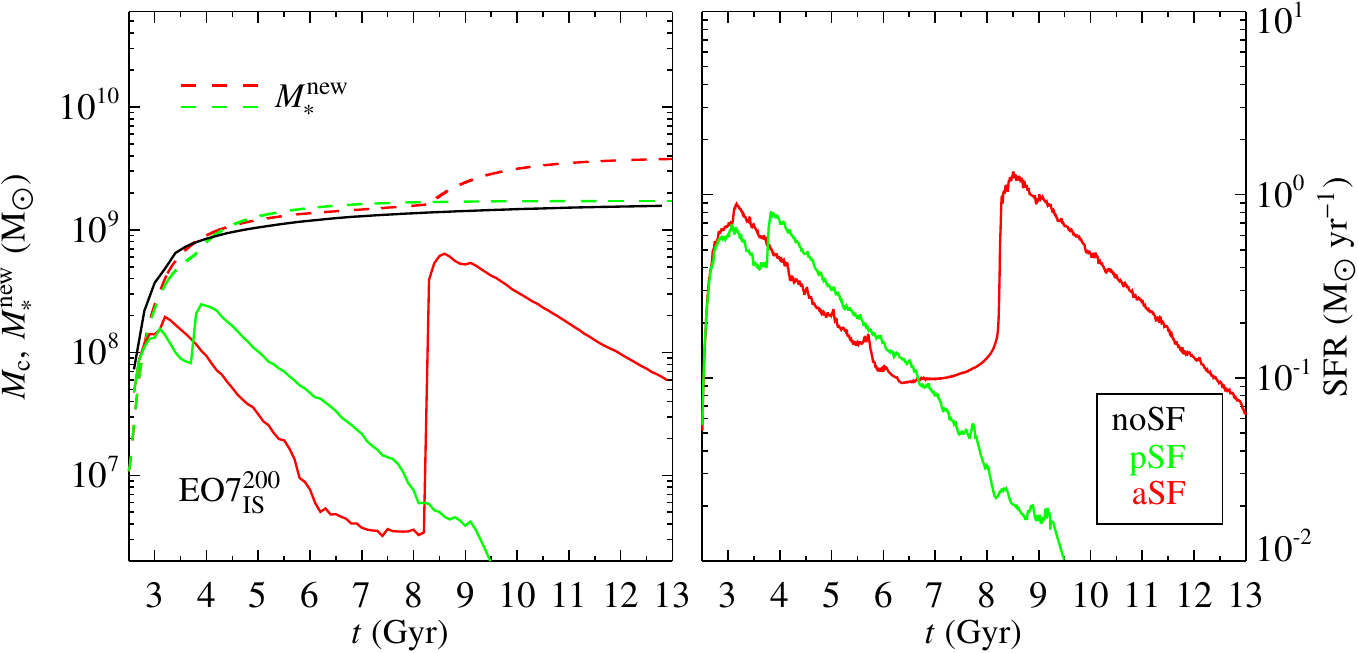}}\\
% 
%\subfloat{\includegraphics[width=0.67916666\linewidth,keepaspectratio]{paper3_mass_NFWi_200_IS_E7_FO}}\\
\subfloat{\includegraphics[width=0.8\linewidth,height=0.35\linewidth]{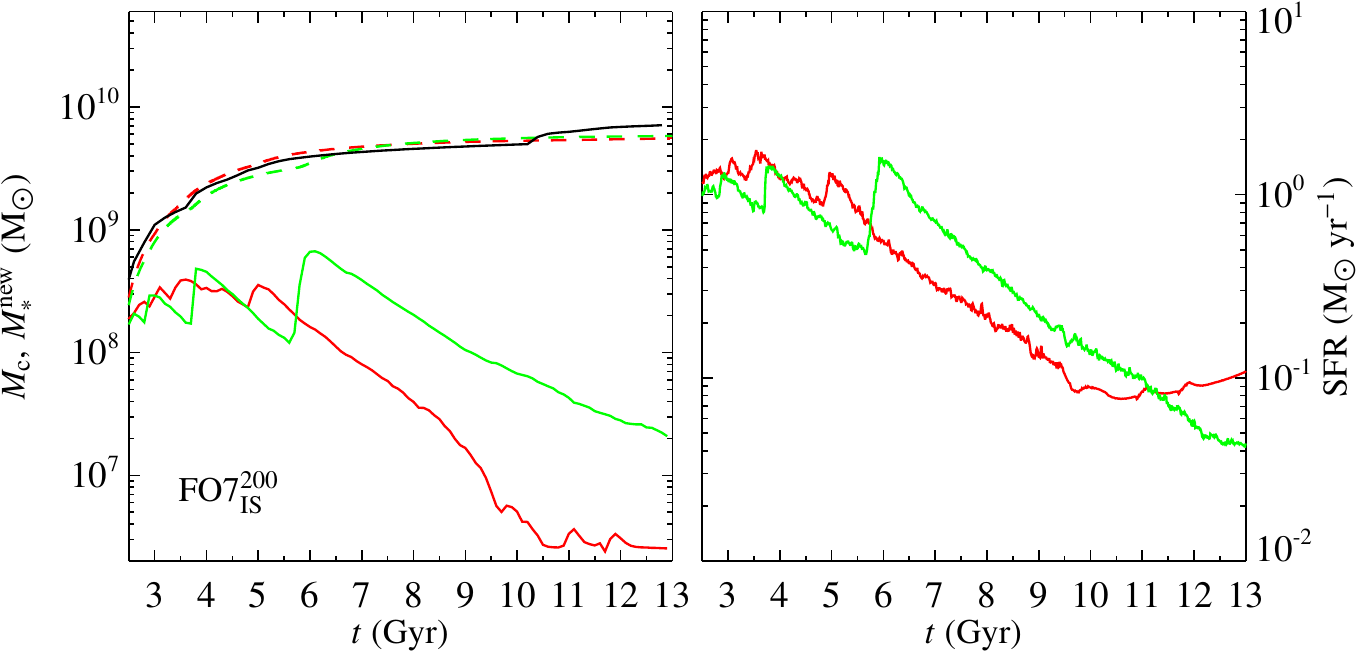}}\\
% 
%\subfloat{\includegraphics[width=0.67916666\linewidth,keepaspectratio]{paper3_mass_NFWi_300_IS_E7_EO}}
\subfloat{\includegraphics[width=0.8\linewidth,height=0.35\linewidth]{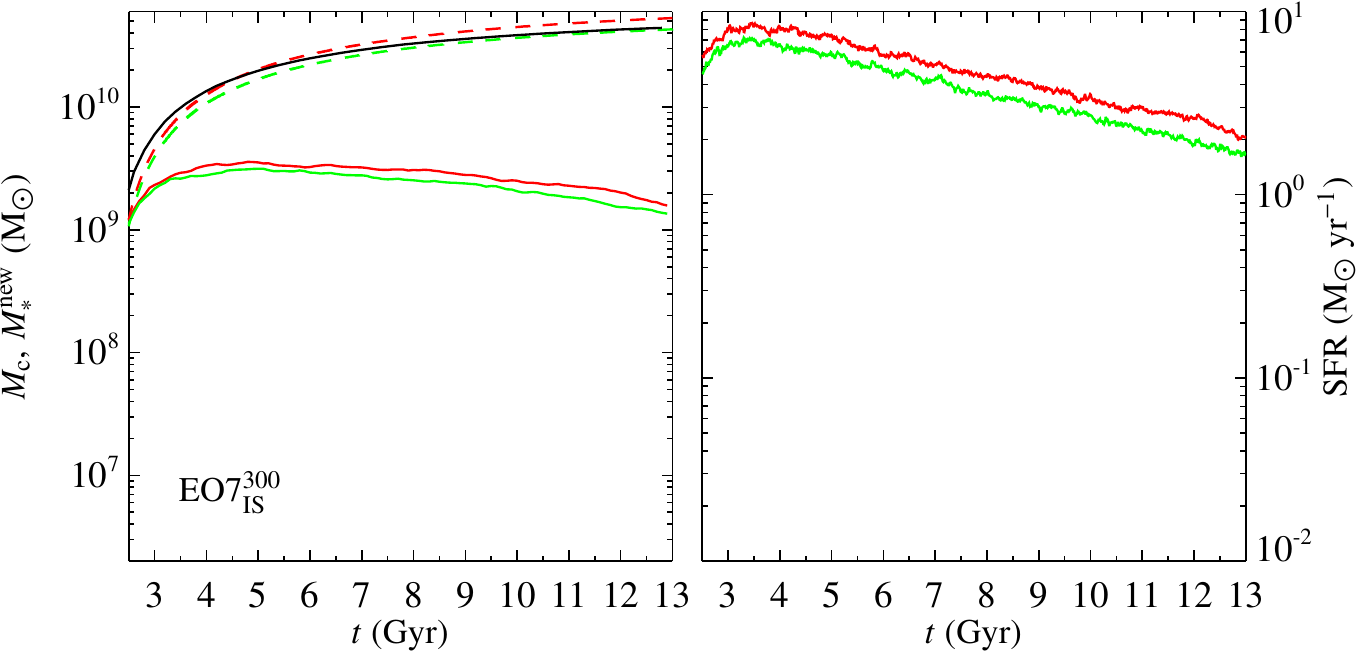}}
\caption{Left panels: time evolution of the cold gas mass $M_\mathrm{c}$ (solid lines), and of the cumulative mass in newly formed stars 
$M_*^{\rm new}$ (dashed lines). Right panels: time evolution of the SFR rates.  The models are the same of Fig.~\ref{LTt}. 
The black line refers to the noSF model,
green and red lines refer to pSF and aSF, respectively, with $\eta_{\rm SF} = 0.1$. Note how $M_\mathrm{c}$ is much reduced 
in the SF models (green and red solid lines) with respect to the noSF models (black lines), and how at the same time $M_*^{\rm new}$
(dashed lines) reaches close to the noSF $M_\mathrm{c}$ values. See Sect.~\ref{mass} for more details.}
\label{Mt}
\end{figure*}

\subsection{Intermediate mass ETGs}\label{interm}

Here we present the results for two representative models, of E4 and
E7 shapes, with intermediate galaxy mass ($\sigma_{\rm e8}=250$ km
s$^{-1}$ for the E0 counterpart, see Tab.~\ref{tab1}), that have been run with
the same kinematic configuration of the LM and HM models presented
previously ($k=1$, isotropic rotator), and also with a lower rotation ($k=0.1$).
For
these, only aSF has been considered, with $\eta_{\rm SF} = 0.01$ and
$\eta_{\rm SF} = 0.1$. The final values for the most relevant quantities
of these eight models are presented in Tabs. \ref{tab4} and \ref{tab5}.

The E4 models with $k=1$ show the same smooth evolution of
the HM models, i.e., no or very small oscillations in $\Lx$ and $\Tx$
(see the dashed lines in Fig.~\ref{s250LT}, similar to
those in Fig.~\ref{LTt}); in fact, the
evolution of the hot gas is similarly not affected by SF. The E7 models with $k=1$,
instead, show oscillations in $\Lx$ and $\Tx$, 
similar to those of the EO7$^{200}_{\rm IS}$ and FO7$^{200}_{\rm IS}$ models, but fewer and less
pronounced (see the solid lines in Fig.~\ref{s250LT}, to be compared
with those in Fig.~\ref{LTt}).
In the aSF model the final $\Lx$ is larger than for the
noSF case, as found for the FO7$^{200}_{\rm IS}$ models (Sect. 3.1).

In models with rotational velocities reduced by a factor of ten
($k=0.1$), $\Lx$ and $\Tx$ increase largely, as expected from the
results of N14 showing a larger $\Lx$ and $\Tx$ in non-rotating
galaxies ($k=0$).  Again the E4 models show no oscillations in $\Lx$
and $\Tx$, while the E7 models show numerous small oscillations in
$\Lx$ and $\Tx$, that increase in amplitude with time increasing
(Fig.~\ref{s250LT}).  The cold gas disc is much less extended: its
radius is $\la 1$ kpc for $k=0.1$, and $\la 10$ kpc for $k=1$
(slightly larger for flatter shape, and for lower $\eta_{\rm SF}$).

\subsection{Mass exchange between the cold disc and the new stars}\label{mass}

The second main result of the new simulations is that most of the cold
gas disc is consumed by star formation, both for pSF and aSF. This result, of general
validity, is shown for the particular models of Fig.~\ref{LTt} in 
Fig.~\ref{Mt}, where the cold gas mass $M_\mathrm{c}$ of the noSF models (black solid line)  
is at the end very close to that of the new stars $M_*^{\rm new}$
(dashed lines). For example, the EO7$_\mathrm{IS}^{\rm 200}$ model in the noSF case has a cold disc extending out to a 2.5 kpc radius, of $M_\mathrm{c}=1.6\times
10^9$M$_{\sun}$, while the corresponding pSF model at the end has no
cold disc, and a mass in new stars of $M_*^{\rm new}\simeq 1.7\times
10^9$M$_{\sun}$  (a bit more gas has cooled in the pSF case than in the
noSF case). Similarly, this same galaxy model, 
with aSF and $\eta_{\rm SF}=0.1$, at the end has a cold disc of just
$M_\mathrm{c} \simeq 5.7\times 10^7$M$_{\sun}$, and $M_*^{\rm new}\simeq
3.8\times 10^9$M$_{\sun}$. 
This $M_*^{\rm new}$ is larger  ($\sim$ double) 
than for the pSF case, a fact that is partly accounted for by
the injected mass from the new stars in the aSF model\footnote{We
  recall that, in the aSF scheme, the mass injected back into the ISM
  by the new stars can itself condense and produce new stars, that in turn
  contribute to the computation of $M_*^{\rm new}$.}, and mostly by the
delay in the final cooling episode after which a galactic wind is
established (Fig.~\ref{LTt}); this delay keeps the galaxy gas-rich and
starforming for a longer time (which is proved also by the lower
escaped mass $M_\mathrm{esc}$ in the aSF than in the pSF case, see Tab.~\ref{tab3}).

\begin{figure*}
%\subfloat{\includegraphics[width=0.75\linewidth,keepaspectratio]{
% Mcold_Mstar_vs_Mstar}}\\
%\subfloat{\includegraphics[width=0.75\linewidth,keepaspectratio]{Mcold_vs_Mstar
% }}\\
%\hskip 4.5truecm
%\subfloat{\includegraphics[width=0.53\linewidth,keepaspectratio]{
% Mstar_new_Mstar_vs_Mstar}}\\
%\subfloat{\includegraphics[width=0.53\linewidth,keepaspectratio]{
% Mstar_new_vs_Mstar}}\\
%\hskip 4.5truecm
%\subfloat{\includegraphics[width=0.53\linewidth,keepaspectratio]{
% SFR_Mstar_vs_Mstar}}\\
%\hskip 4.5truecm
%\subfloat{\includegraphics[width=0.53\linewidth,keepaspectratio]{
% mean_time_vs_Mstar}}
\subfloat{\includegraphics[width=0.73\linewidth,keepaspectratio]{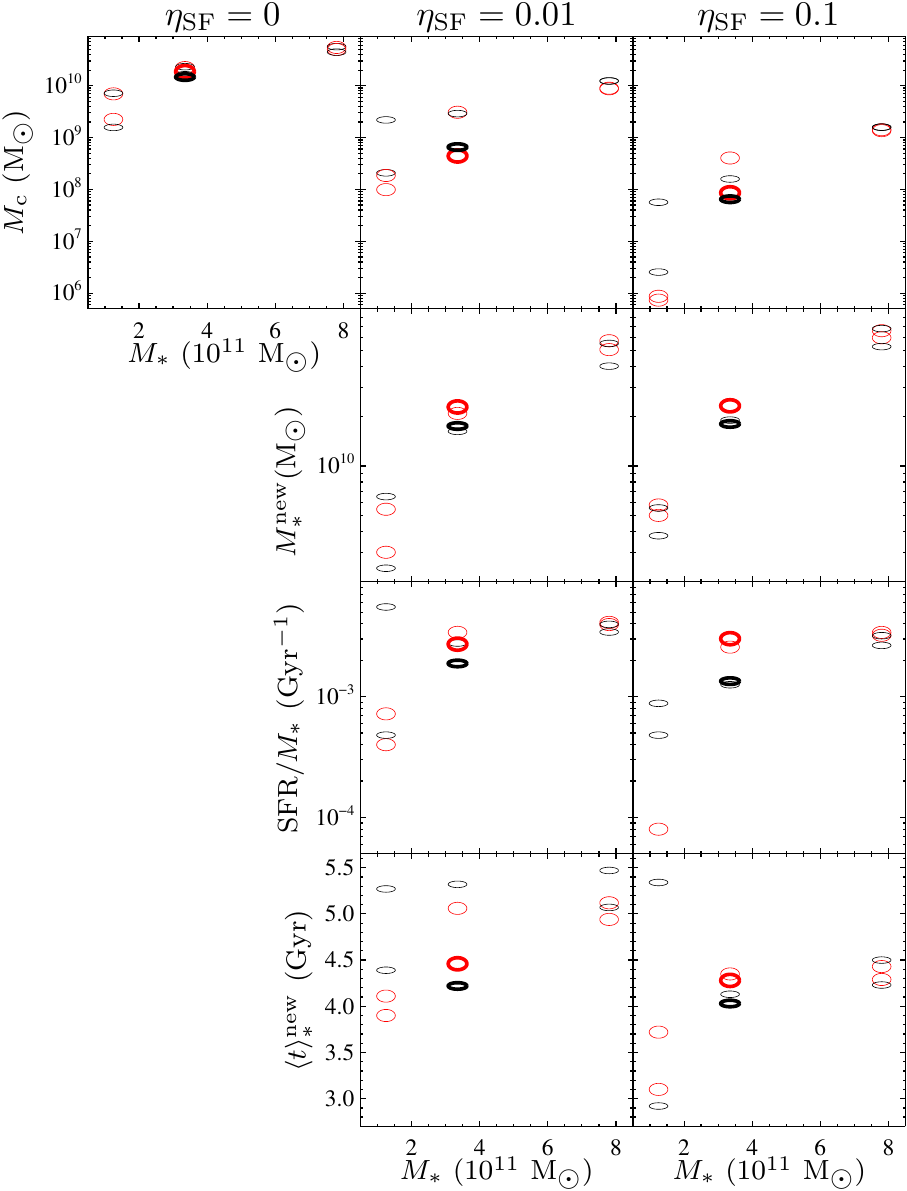}}
\caption{Top panels: final values of the cold gas mass, 
%normalized to the original stellar mass of the galaxy $M_*$,
for the models in Tabs. \ref{tab2}-\ref{tab5} without SF (left panel), and 
with aSF,  for the two different $\eta_{\rm SF}$ values adopted here.
Middle and bottom panels: final values of the stellar mass in newly formed stars $M_*^{\rm new}$, of the SFR
normalized to the original $M_*$, 
and of the mean formation time of the new stars $\langle t\rangle _*^{\rm new}$, 
calculated from the beginning of the simulation, 
for the same models with aSF in the top panels. All values are given in Tabs.
\ref{tab2}-\ref{tab5}. See Sects.~\ref{mass}
and~\ref{hist} for more details.}
\label{Mcold}
\end{figure*}

In general, the LM models with SF show a final $M_\mathrm{c}$ that can range
from 0 to $10^9$M$_{\sun}$, but is typically $\la 10^8$M$_{\sun}$
(Fig.~\ref{Mcold} and Tab.~\ref{tab3}).  The final mass in the newly formed
stellar disc is of the order of $M_*^{\rm new}\simeq $a few$\times 10^9$~M$_{\sun}$; thus,
the $M_*^{\rm new}$ values reach almost those of the $M_\mathrm{c}$ of the
corresponding
noSF models (Tab.~\ref{tab3}; Fig.~\ref{Mcold}).
%Figure~\ref{Mt} shows that most of SF comes from $M_\mathrm{c}$, i.e., gas at a
temperature below 2$\times 10^4$K. 

Also in the HM models, similarly to what found for the LM models,  $M_*^{\rm new}$ 
at the end of the simulations 
is of the order of the $M_\mathrm{c}$ values of the noSF models.
A larger massive disc remains in the HM models, though, even when including SF
(e.g., Fig.~\ref{Mt} and~\ref{Mcold}).  Their final $M_\mathrm{c}$ is $\simeq
10^9$ M$_{\sun}$, for
$\eta _{\rm SF}=0.1$ ($\simeq 1/50$ of $M_\mathrm{c}$ without SF), and ($8-12)\times
10^9$ M$_{\sun}$, if $\eta
_{\rm SF}=0.01$ ($M_{c}\simeq 1/5$ of $M_\mathrm{c}$ without SF), independently
of pSF or aSF.  
As found for the LM galaxies, in the aSF models
$M_*^{\rm new}$ is larger than in the pSF ones, and can be even larger than
$M_\mathrm{c}$ of the noSF models (Figs.~\ref{Mt} and~\ref{Mcold}).  This time, though, the
difference is lower, of $\simeq 20$\%, and corresponds to secondary SF
sustained by the material injected by the new stars. The
final mass in the newly formed stellar disc is of the order of $M_*^{\rm new}\simeq $a
few$\times 10^{10}$ M$_{\sun}$ (Tab.~\ref{tab2}).  However, it must be noted
that the structural properties of the HM models may not be very
realistic: ETGs of this mass with the isotropic rotator kinematics are
quite unlikely, since they are not observed to rotate this much
\citep{emsellem.etal2011}. Real massive galaxies will have less rotation
than adopted here with $k=1$, and then they will have less cold
gas and less new stars. 

More realistic massive and rotating ETGs are those of Sect.~\ref{interm}, with aSF.  For
these, if $k=1$, the final cold gas mass is again largely reduced when
introducing SF, by a factor of $\simeq 50$ (E4) and $\simeq 100$ (E7),
when $\eta_{\rm SF} = 0.1$. At 13 Gyr, $M_\mathrm{c}$ is few$\times
10^9$M$_{\sun}$ for $\eta_{\rm SF}=0.01$, and few$\times
10^8$M$_{\sun}$ for $\eta_{\rm SF}=0.1$; $M_*^{\rm new}$ is $\simeq
2\times 10^{10}$M$_{\sun}$, for both $\eta_{\rm SF}$ values.
If $k=0.1$, the final $M_\mathrm{c}$ is reduced even more than for the $k=1$ models 
(Fig.~\ref{Mcold}), for both the E4 and E7 shapes:
$M_\mathrm{c}$ is a few $\times 10^8$M$_{\sun}$ ($\eta_{\rm SF} = 0.01$), and $\la 10^8$M$_{\sun}$ ($\eta_{\rm SF} = 0.1$).
Note that the final cold mass $M_\mathrm{c}$, without SF, is
similarly large for the $k=1$ and $k=0.1$ models (Tabs. \ref{tab4} and
\ref{tab5}; Fig.~\ref{Mcold}), but 
it is instead lower for the $k=0.1$ models than for the $k=1$ ones, when adding SF;
one could conclude that  in the $k=0.1$ models SF is more efficient, because it takes place 
in a cold disc which is smaller and denser, or that the evolution is different, if less gas 
overall cooled (the heating was more efficient). The second hypothesis is supported by the finding 
that, at the end of the simulations, $M_*^{\rm new}$ is similar for $k=1$ and $k=0.1$
(e.g., $2.4$ and $2.3\times 10^{10}$M$_{\sun}$ for the E4 models, respectively with $k=1$ and $k=0.1$,
for $\eta_{\rm SF}=0.1$), and similarly larger 
than $M_\mathrm{c}$ of the corresponding noSF models (e.g., 
$2.0$ and $1.9\times 10^{10}$M$_{\sun}$, for the E4 models respectively with
$k=1$ and $k=0.1$).

In conclusion, SF seems to be an important mechanism to solve the worrysome feature of massive cold 
gas discs in rotating ETGs without SF.
In addition, we found the following trends, at the end of the simulations (see also 
the top two rows of panels of Fig.~\ref{Mcold}):
$M_\mathrm{c}$ increases with $M_*$, and has a large spread for LM models, that increases for larger $\eta_{\rm SF}$;
$M_\mathrm{c}/M_*$ is roughly constant with $M_*$, with a spread reaching down to values much lower 
than this constant for LM models; at any $M_*$, $M_\mathrm{c}$ decreases when $\eta_{\rm SF}$ increases.
$M_*^{\rm new}$ and $M_*^{\rm new}/M_*$ increase with $M_*$, with a spread for LM models, and are 
roughly independent of $\eta_{\rm SF}$. This latter result means that more massive ETGs have been overall
more efficient in forming stars from recycled material, considering their past $\simeq 11$ Gyr.

\subsection{The SF history}\label{hist}

SF is not only responsible for the final mass budget of $M_\mathrm{c}$ and $M_*^{\rm new}$, but also
for the rate of SF (SFR), its evolution, and the (fiducial) age of $M_*^{\rm
new}$;
that is, for the SF history that we discuss here.
Star formation shows a different evolution in LM and HM galaxies, in
pace with their different ISM evolution \referee{(see Fig.~\ref{Mt})}.  In LM galaxies, where major
cooling episodes take place recursively, the SFR peaks
suddenly in correspondence of these cooling events \referee{due to the increased density of cold gas}, and declines right
after, going eventually to practically zero \referee{either} if \referee{the cold gas reservoir is completely consumed or} the cooling episode is
followed by a global wind (Sect.~\ref{comp}).  Thus, in general LM
galaxies have a larger and peaked SFR in their past (reaching 1-2
M$_{\sun}$ yr$^{-1}$), when the rate of stellar mass losses due to
the original stellar population was much larger, and a low SFR ($\la
0.1$M$_{\sun}$ yr$^{-1}$) at the present epoch.  Typical values at 13
Gyr can be as low as zero, and at most as high as 0.45 M$_{\sun}$
yr$^{-1}$ (Tab.~\ref{tab3}). These values compare well with the current rates
estimated for the ATLAS$^{\rm 3D}$ sample, that range from $\approx 0.01$ and 3
M$_{\sun}$ yr$^{-1}$,
with a median value of 0.15
M$_{\sun}$ yr$^{-1}$ for ETGs of $\sigma_{\rm e8}$ comparable to
those of the LM models run here \citep{davis.etal2014}.  HM galaxies,
instead, show a more regular and steady production of cold gas, and so
is their SFR (Fig.~\ref{Mt}); at 13 Gyr, their SFR is 2-3 M$_{\sun}$
yr$^{-1}$ (Tab.~\ref{tab2}; but we recall these models are not realistic). In
the intermediate mass models (Sect.~\ref{interm}), the SFR is larger
for the E4 models without oscillations in $\Lx$ than for the E7 ones.
Overall, the SFR at 13 Gyr is similar for $k=1$ and $k=0.1$: for
$k=1$, the SFR ranges from 0.42 to 1.1 M$_{\sun}$ yr$^{-1}$ (i.e., it
is larger than for the LM models of same $k=1$); when $k=0.1$, the SFR
ranges from 0.45 to 1.0 M$_{\sun}$ yr$^{-1}$.
Figure~\ref{Mcold} summarizes the {\it present epoch} SFR behaviour, as a function of 
$M_*$,  for the aSF models.
To compare models of different mass, the plotted quantity is the SFR normalized to $M_*$.
This quantity is
$\sim (1-4)\times 10^{-3}$~Gyr$^{-1}$, quite independent of $\eta_{\rm SF}$, for
the intermediate mass and HM models; it can be largely varying for the
LM models (from $4\times 10^{-4}$ to $6\times 10^{-3}$~Gyr$^{-1}$, for
$\eta_{\rm SF}=0.1$, and from $8\times 10^{-5}$ to $\sim 10^{-3}$~Gyr$^{-1}$, for $\eta_{\rm SF}=0.01$).  On average, the present epoch
SFR/$M_*$ increases with $M_*$, a result similar to that
of the previous Sect.~\ref{mass}, where we found that the integration over time of the SFR, 
that is $M_*^{\rm new}$,
and $M_*^{\rm new}/M_*$, increase with $M_*$.

Tables \ref{tab2} and \ref{tab3} also list the mean formation time $\langle t\rangle _*^{\rm new}$
of the new stars since the beginning of
the simulation ($t_0=2$ Gyr), and the mean star formation rate (defined as $M_*^{\rm new}/\langle t
\rangle _*^{\rm new}$),
both at 13 Gyr.
$\langle t\rangle _*^{\rm new}$ as  a function of time is defined as:
\begin{equation}
\langle t\rangle _*^{\rm new} (t)=\dfrac{1}{M_*^{\rm new}(t)} \int _{t_0} ^t (t^{\prime}-t_0) \, {\rm SFR}(t^{\prime}) \diff t^{\prime},
\end{equation}
where the SFR$(t)$ is the volume integrated, instantaneous star formation rate.
The values of $\langle t\rangle _*^{\rm new}$ at the  {\it present epoch} are shown in Fig.~\ref{Mcold} 
for the aSF cases. They range from 2.4 to 6.6 Gyr, for the LM models,
with a  tendency to be lower for the aSF than for the pSF, and for the larger $\eta_\mathrm{SF}$.
The HM and intermediate mass models have $\langle t\rangle _*^{\rm new}$ in a narrower range (from 4.2 to 5.5 Gyr),
independent of the aSF or pSF, and are again lower for the larger $\eta_{\rm SF}$ (Fig.~\ref{Mcold}). 
Therefore, the new stars
in LM models have a larger spread of formation times,
and correspondingly a larger range of ages at the current epoch, 
than in larger mass models. 

For the aSF, for $\eta_\mathrm{SF}=0.01$, the upper envelope of the more recent  
formation times (largest $\langle t\rangle _*^{\rm new}$ values) is $\simeq 5.5$ Gyr,
roughly independent of $M_*$, while
the lower envelope (lowest $\langle t\rangle _*^{\rm new}$ values) increases with $M_*$.
A similar trend is shown by the HM and intermediate mass models, 
for $\eta_{\rm SF}=0.1$, 
but shifted towards lower $\langle t\rangle _*^{\rm new}$ values ($\la 4.5$ Gyr);
the spread of $\langle t\rangle _*^{\rm new}$ for LM models keeps instead large, 
extending to even lower $\langle t\rangle _*^{\rm new}$ values of $\simeq 3$ Gyr.
The decrease of $\langle t\rangle _*^{\rm new}$ for larger $\eta_{\rm SF}$, i.e., for a more efficient SF,
obtained here from hydrodynamical simulations provides support to the similar conclusion 
drawn in galaxy formation models focussed on chemical evolution: these 
assume that the efficiency of star formation is an increasing function of mass, which for them
has -- among others -- the consequence that smaller galaxies continue to form stars for longer periods, while more
massive galaxies host an older stellar population \citep[e.g.,][]{matteucci1994}.

It is interesting that $\langle t\rangle _*^{\rm new}$ decreases slightly, or remains similar,
when switching from pSF to aSF; in fact, one could expect a more extended 
SF, due to the injection of material from the new stars, for the aSF. 
Evidently, the net results of the injection of both mass 
and energy is to produce the bulk of 
SF at roughly the same time, for the pSF and aSF schemes; only, the final mass in new stars is larger for aSF.

The mean age of the newly formed stars in Gyr is 13$-t_0-\langle t\rangle _*^{\rm new}$. 
This age is in any case larger than $\sim 5.5$ Gyr, for the aSF.

\begin{figure*}
\includegraphics[width=0.65\linewidth,keepaspectratio]{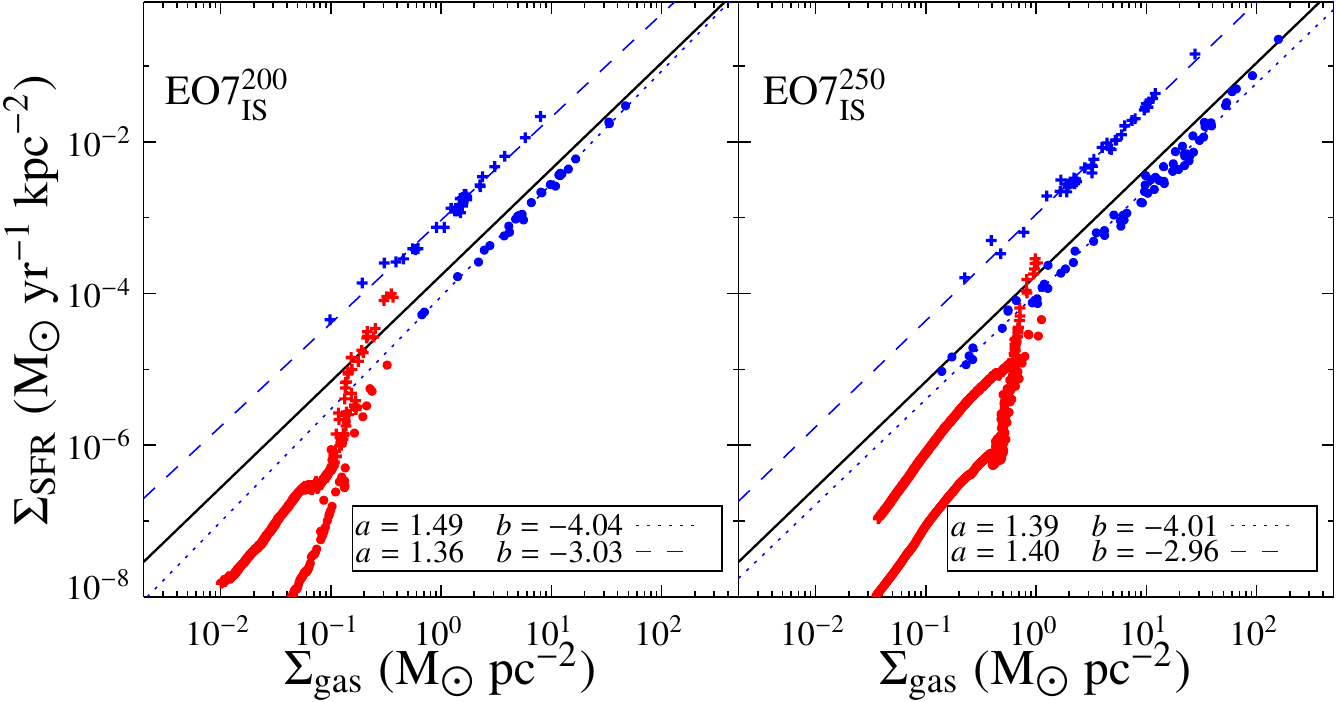}
%\subfloat{\includegraphics[width=0.45\linewidth,keepaspectratio]{kenn3_2007EO}}
% 
%\subfloat{\includegraphics[width=0.45\linewidth,keepaspectratio]{kenn3_2507EO}}
\caption{Projected SFR versus mass surface density of
the gas (at $T<2\times 10^4$K, for the blue symbols, and at $T>2\times 10^4$K,
for the red ones), for two aSF models: EO7$^{200}_{\rm IS}$ on the left, and
EO7$^{250}_{\rm IS}$ on the right.
Full  circles refer to $\eta_\mathrm{SF}=0.01$ and crosses to $\eta_\mathrm{SF}=0.1$. 
The lower $\eta_{\rm SF}=0.01$ coefficient for the SF
recipe (eq.~\ref{eq:sf}) reproduces better the empirical
Kennicutt-Schmidt scaling, that is shown by the solid line (with a
slope of $N=1.4$, and normalized as in \citet{davis.etal2014} for their ATLAS$^{\rm 3D}$ sample
of molecular gas-rich ETGs). The blue
dotted and dashed lines show the best fits for the cold gas only
(coefficients in the frames in the lower right corners). See
Sect.~\ref{KS} for more details.}
\label{KSfig}
\end{figure*}
\subsection{The SF recipe and the Kennicutt-Schmidt relation}\label{KS}
From eq.~\ref{eq:sf} it follows that SF is favoured in regions of high density, and relatively short $t_\mathrm{SF}$.
These conditions are naturally present in low temperature regions; accordingly, the simulations show that 
most of the SF  takes place where the gas
temperature is $\la 2\times 10^4$K, i.e., in the cold rotating disc.

Quite obviously, our results depend somehow on the adopted prescription 
for the SF, while a realistic, physically consistent,
description of SF is not available. Yet, it is interesting to ask at what level 
our implementation of SF recovers some basic observational features of SF.
More than half a century ago \citet{schmidt1959} conjectured that the rate
of SF should vary as a power law of the gas density, and later
\citet{kennicutt1998} suggested the parameterization in terms of mass surface densities $\Sigma_{\rm SFR} =
A\Sigma_\mathrm{cold\, gas}^N$, valid for
 starforming galaxies, with $N\simeq 1.4-1.5$ \citep{kennicutt.evans2012}. 
 We investigate here what scaling (if any) the
adopted SF recipe of eq.~\ref{eq:sf} produces in terms of $\Sigma_{\rm SFR}$
vs. $\Sigma_{\rm gas}$, when implemented in the simulations.
%Note that the only additional ingredient in the adopted prescription  for  SF in eq.~\ref{eq:sf} is the maximum between $t_{cool}$ and 
%$t_{dyn}$ for the gas.

Figure~\ref{KSfig} shows for a few representative models at the end of
the simulation the relationship between $\Sigma_{\rm SFR}$ and
$\Sigma_{\rm gas}$, calculated separately for the hotter (T$> 2\times
10^4$K) and colder (T$\leq 2\times 10^4$K) phases of the gas;
$\Sigma_{\rm SFR}$ is the face-on projection of $\dot\rho_\mathrm{SF}$.  The
figure shows for comparison a recent scaling for the Kennicutt-Schmidt
\referee{relation}, derived by \citet{davis.etal2014} for the ATLAS$^{\rm 3D}$ sample of
molecular gas-rich ETGs (this has a slope of $N=1.4$).
Figure~\ref{KSfig} tells that SF in the gas takes place basically in
two regimes, and by far most of it is due to cold gas.  The behaviour
of the cold gas shows two remarkable features: the first is that a
scaling similar to that of the Kennicutt-Schmidt \referee{relation} is followed quite
closely by the blue (cold gas) points; the second is that even the
normalization is reasonably well reproduced by our scheme of ``active
SF'', when $\eta_{\rm SF}=0.01$ (models with $\eta_{\rm SF}=0.1$
correspond instead to a normalization that is too large).  In
particular, the cold gas in the log$\Sigma_{\rm SFR}$--log$\Sigma_{\rm
  gas}$ plane follows a trend best-fitted by a line of slope of $\simeq
1.4-1.5$, a fact that can be explained as follows.  SF in the cold gas
is most likely regulated by the dynamical timescale $t_\mathrm{dyn}$, being
$t_\mathrm{cool}$ very short at such temperatures and densities; thus, one
roughly expects that $\dot\rho_{\rm SF} \propto \rho^{1.5}$. A slope
for the $\Sigma_{\rm SFR}$--$\Sigma_{\rm gas}$ relation close to the
empirical one then follows when neglecting the difference between
volume and surface density, which is not unreasonable considering that
the cold disc resides mainly in a thin region above and below the
equatorial plane, extending vertically for just a couple of gridpoints
($< 200$ pc; see also  \citealt{kennicutt.evans2012}).  For the hotter gas,
the red points in Fig.~\ref{KSfig} are invariably fitted by a steeper
line, with slope $\simeq 2$; this is explained by a longer cooling
time than the dynamical time, and then by $\dot\rho_{\rm SF} \propto
\rho^{2}$.

Overall, Fig.~\ref{KSfig} proves that the recipe of eq.~\ref{eq:sf} is
very reasonable for SF, and, together with the chosen range of
normalization for $\eta_{\rm SF}$, it provides a scaling close to that
observed for normal to starburst galaxies.  These considerations add
strength to the overall results obtained in this study.
\section{Conclusions}

In this work we have explored the effects of SF on the evolution of
rotating hot gas flows in early-type galaxies; our previous (N14) high resolution 2D hydrodynamical
simulations of such flows, run for axisymmetric two-component
models, were successful in producing the observed $\Lx$ and $\Tx$ 
of flat and rotating ETGs, but also revealed the formation of  very massive
cooled gas discs. To study the SF effects, we performed
hydrodynamical simulations using the same numerical code of N14, where
SF with a Salpeter IMF is inserted, following a simple recipe
depending on the gas density, and the cooling and dynamical timescales of the ISM
(eq.~\ref{eq:sf}), with two possible values for the efficiency of SF
($\eta_{\rm SF}=0.01$ and 0.1).  We considered the new stellar generations
to be passive (a pure sink of gas), or active (the new stars
contribute mass and energy to the gas flow, via stellar winds and
SNIIs). The new simulations have been run for 
a subsample of the N14 models, of high, intermediate and low galaxy masses, of E4 and E7 shapes,
with the ordered stellar rotation described by the isotropic rotator, or 
with the rotational velocities reduced by a factor of ten.

It is found that subsequent generations of stars are formed from the
cold gas that accumulates in the equatorial
plane, and that most of the extended and massive cold disc found by N14 is consumed
by this process. In particular, the main conclusions are summarized as follows.

$\bullet $  Remarkably, both for the passive and the active SF
implementation, we confirm the results of the previous investigation
without SF, concerning the trends of $\Lx$ and $\Tx$ of the hot ISM with
galactic rotation. As found by N14,  $\Lx$ is lowered by rotation: in
low mass ETGs this happens because rotation favours the establishment of global winds,
and in medium-to-high mass ETGs because rotation lowers the hot gas density 
in the central galactic region. The average temperature $\Tx$ is also
lower in intermediate/high mass ETGs, when rotation is important, due
to the lower contribution of the central regions (usually hotter and
denser in non-rotating systems), while it can be higher in low mass ETGs, if rotation triggers
a wind, due to the thermalization of meridional winds. 

$\bullet $ The robustness of the N14 results even after the addition of SF is explained differently for HM and LM galaxies.
In HM galaxies the evolution of the hot ISM is only marginally sensitive
to the presence of SF, and $\Lx$ and $\Tx$  evolve smoothly as without SF: a
massive cold disc forms early, it is mostly consumed by SF, and around
it the hot ISM evolution proceeds almost unaltered by SF (even in
presence of energy and mass injection by SNII’s).  In less massive
models, instead, the evolution of $\Lx$ is not smooth, but shows many peaks
corresponding to major cooling episodes, as in N14. The time-occurrence of the
peaks depends on SF, thus SF can induce variations of $\Lx$ at a
certain chosen epoch; however, these keep within the already large range of $\Lx$ values
typical of these masses, even without SF, being the gas flow very sensitive 
to many factors \citep[as already found previously, e.g.,][N14]{ciottietal1991}.

$\bullet $ For what concerns the global gas mass budget, at the end of the simulations the cold gas mass left in the equatorial disc
$M_\mathrm{c}$ on average increases with $M_*$, with a larger spread for LM models; at fixed $M_*$,
$M_\mathrm{c}$ is lower for lower stellar ordered rotation, and for larger  $\eta_{\rm SF}$. In addition, the ratio
$M_\mathrm{c}/M_*$ is roughly constant with $M_*$, with a spread reaching down to values much lower 
than this constant for LM models.
In any case, $M_\mathrm{c}\la 2\times 10^9$~M$_{\sun}$ for galaxies with $M_*\leq 4\times 10^{11}$~M$_{\sun}$.
Typical values for $M_\mathrm{c}$ are $\la $a few $\times
10^7$~M$_{\sun}$ for LM galaxies (with the maximum rotational level of
$k=1$), with exceptions of $M_\mathrm{c}\simeq 10^9$~M$_{\sun}$; and a few
$\times 10^7$~M$_{\sun}$ (if $k=0.1$) to $3\times 10^9$~M$_{\sun}$ (if
$k=1$) for ETGs of intermediate mass. These values compare well with
those recently observed \citep{young.etal2014, serra.etal2012, serra.etal2014}. 
In particular, they can explain the observation that 
massive, fast-rotating ETGs often have kinematically aligned gas, independent
of environment \citep{davis.etal2011, davis.etal2013}.

$\bullet $ The mass in newly formed stars $M_*^{\rm new}$, and the ratio $M_*^{\rm new}/M_*$, 
increase with $M_*$, again with a large spread for LM models, and are 
roughly independent of $\eta_{\rm SF}$. This result means that more massive (rotating) ETGs have been overall
more efficient in forming stars via recycling of their stellar mass losses, 
during the past $\simeq 10$ Gyr.
The mass in secondary generations of stars is $(1-6)\times
10^9$~M$_{\sun}$ (if $k=1$) for LM models, and $\simeq 2\times
10^{10}$~M$_{\sun}$ (for both $k=1$ and $k=0.1$) for 
intermediate mass models. These should reside mostly in a disc, as most fast rotator ETGs possess, 
and be related to the birth of a younger, more metal rich disky stellar component that is indeed 
observed \citep{krajnovic.etal2008, cappellari.etal2013}.
They should not be recognized as a ``young'' stellar population in an absolute sense, though, 
since most of $M_*^{\rm new}$ formed a few Gyr ago, as can be evaluated from the
evolution of their SFR.

$\bullet $ The time evolution of the SFR depends on the mass of the galaxy.
LM galaxies ($M_*=1.25\times 10^{11}$~M$_{\sun}$) have a larger and
peaked SFR in their far past (reaching 1-2~M$_{\sun}$ yr$^{-1}$),
when the rate of stellar mass losses due to the original stellar
population was much larger, and when major cooling episodes were frequent. In
fact, the average age of the 
new stars of LM models ranges from 5 to 8 Gyr (for the aSF scheme).  At the present epoch, their
SFR is low ($\la 0.1$~M$_{\sun}$ yr$^{-1}$ typically), with the full
range of values going from zero to $\simeq 0.7$~M$_{\sun}$
yr$^{-1}$. These results agree nicely with the low degree of SF and young
stellar populations that is detected only in fast rotators, in the
ATLAS$^{\rm 3D}$ sample \citep{kuntschner.etal2010, sarzi.etal.2013}; also, they
compare well with the current estimated rates, whose median value is
$\approx 0.15$~M$_{\sun}$ yr$^{-1}$ \citep{davis.etal2014}.
More massive galaxies, instead,
show a more regular and steady production of cold gas, and so is their
SFR;  the average age of their new stars 
ranges from 5.5 to 7.0 Gyr (in the aSF scheme).
At 13 Gyr the SFR is $\simeq (0.4-1)$~M$_{\sun}$ yr$^{-1}$, 
larger than for LM models, and somewhat larger than observed
\citep[e.g.,][]{mcDermid2015}; this could be fixed by
assuming
a lower $\eta_{\rm SF}$, or a lower ordered stellar rotation,
at larger galaxy masses (Fig.~\ref{Mcold}). Thus,  our models may  still be
rotating too much and producing too much cold gas, at galaxy
masses $>2\times 10^{11}$~M$_{\sun}$.  
The current SFR/$M_*$, instead, remains quite constant from LM
to intermediate mass models, with a larger spread for LM models (as already found for other properties).

Finally, the SF recipe adopted for this
work proved to be a reasonable one, given that it reproduces the slope of the
Kennicutt-Schmidt \referee{relation}, and even the normalization if $\eta_{\rm SF}\simeq 0.01$. 

Overall, we can suggest an origin (mostly) in the SF from cooling hot gas,
for the presence of cold gas phases
kinematically aligned with the stars, and for the low-level degree of SF, all features detected only in fast
rotators in the ATLAS$^{\rm 3D}$ sample.

In the perspective of the galaxy evolution, we finally recall that 
the precise knowledge of the amount of gas flowing towards the galactic centre is of
great importance for a proper study of feedback effects from the
central black hole in rotating galaxies \citep[e.g.,][]{novak.etal2011, gan.etal2014}.  A major 
implication of the present work is the fact that, if the gas produced
by stellar evolution forms new stars, the amount of gas in principle available for
fuelling the central supermassive black hole is much lowered. Of course, such fuelling would be 
impossible in presence of rotation and absence of viscosity. 
Disc
instabilities, though,  could break axial symmetry, and allow for
the gas infall onto the black hole, possibly escaping SF.
Phenomenologically, the effect of gravitational instabilities in the disc
can be modelled as an effective gravitational viscosity \citep[e.g.,][]{bertin.lodato.2001, hopkins.quataert2011, rafikov2015}, that favours accretion of cold
gas towards the centre.  Indeed, viscosity (that could be due for example to MRI)
is another ingredient not present in the current simulations, but that could affect the disc
evolution.  As the code is axisymmetric, we cannot follow the
complex physics of (non-axisymmetric) disc instabilities that could be
present.
Although self-gravity of the gas is not considered in our simulations, yet
for some models at a selection of times we computed the fiducial
value of the radial profile of the stability
parameter $Q(R)= c_\mathrm{s} k_R/\upi G\Sigma_\mathrm{c}$, where $k_R$ is the local
epicyclic frequency of the galaxy, and $c_s$ and $\Sigma_\mathrm{c}$ are
respectively the vertically mass-averaged values of the ISM sound
velocity, and the surface density of the cold rotating gas. The $Q$ values were computed
for a layer $\Delta z\simeq 200$ pc thick above and below the galaxy equatorial
plane; these $Q$  turned out to be quite independent of the
adopted thickness value $\Delta z$, being dominated by the cold disc. 
These fiducial $Q$ values maintained
invariably larger than unity (indicating stability) when SF is
allowed, while $Q <1$ on a central $\simeq$kpc-size region, in absence of SF. From this preliminary
and qualitative analysis, it follows that SF {\it and} $Q$-instability are both present in
the gaseous rotating discs. The relevant question is then if and what
of the two processes is dominant in the disc. In a future work we plan to extend
further the present investigation by implementing in the code
the cooperating effects of SF and mass discharge
from the disc to the centre. In this new study, we will also consider
the neglected late-time mass return from the newly born stellar
population, that is the mass injected by stars of mass
$<8$~M$_{\sun }$; this input could be modelled with the method described
in \citet{calura.etal2014}.
\section*{Acknowledgements}
We acknowledge Giuseppe Lodato and Jerry Ostriker for useful discussions, and Silvia Posacki for 
providing the galaxy models of N14. 
L.C. and S.P. were supported by the MIUR grant PRIN 2010-2011, project
`The Chemical and Dynamical Evolution of the Milky Way and Local Group
Galaxies', prot. 2010LY5N2T.

%\end*{acknowledgements}

% \bibliographystyle{mn2e}
% \bibliography{citations}

%%%%%%% Tables:

% \newpage
\clearpage

\renewcommand\arraystretch{1.3}
\setlength{\tabcolsep}{5pt}
\begin{sidewaystable*}
%\begin{table*}
%\centering
 \vskip 15truecm
 \hskip 1truecm
\caption{$\,\,\,\,$ Relevant quantities at $t=13$Gyr for the rotating models
($k=1$) with $\sigma_{\rm e8}$=300 km s$^{-1}$.} 
%\scriptsize
\small
\begin{tabular}{ccccccccccccccccc}
\toprule
name                    &     SF      & $\eta_\mathrm{SF}$&     $M_\mathrm{inj}$  &   
$M_\mathrm{esc}$   &$M_\mathrm{gas}$    &    $M_\mathrm{hot}$   &   $M_\mathrm{c}$&       $M_*^{\rm
new}$&        $M_\mathrm{inj}^{\rm II}$&  $\langle t\rangle _*^{\rm new}$ & 
$M_*^{\rm ew}/\langle t\rangle _*^{\rm new}$ &        $\mathrm{SFR}$ & $\Lx$ & $L_{\rm
SNIa}$ & $L_{\rm SNII}$ & $\Tx$ \\
                       &             &                   & $(10^9$M$_{\sun})$   
 &  $(10^9$M$_{\sun})$     & $(10^9$M$_{\sun})$    & $(10^9$M$_{\sun})$   &  
$(10^9$M$_{\sun})$    & $(10^9$M$_{\sun})$  &  $(10^9$M$_{\sun})$    & (Gyr)    
   & (M$_{\sun}$ yr$^{-1}$)  & (M$_{\sun}$ yr$^{-1}$)   & $(10^{40}$erg
s$^{-1})$ & $(10^{40}$erg s$^{-1})$  & $(10^{40}$erg s$^{-1})$  & (keV) \\
(1)                     &   (2)       &          (3)      &    (4)       &    
(5)      &  (6)         &     (7)      &    (8)       &   (9)        &       
(10)  &  (11)        & (12)         & (13)     & (14) & (15) & (16) & (17)    
\\
\hline
EO4$^{300}_{\rm{IS}}$    & $\times$    &   --              &        69.38 &     
  14.10 &        56.04 &         6.52 &     49.8     &      --   &         --  
&          --   & --   &     --  & 1.34   & 47.7  & -- & 0.68 \\    
EO7$^{300}_{\rm{IS}}$    & $\times$    &   --              &        65.46 &     
  15.87 &        50.24 &         6.13 &     44.5     &       --   &         --  
&          --   & --    &    --  & 1.04  & 45.0 & -- &  0.56 \\    
FO4$^{300}_{\rm{IS}}$    & $\times$    &   --               &        71.81 &    
   11.93 &        60.92 &         6.11 &     54.7     &       --   &         -- 
 &          --   & --   &     --  & 1.35  & 49.4 & -- &  0.68  \\    
FO7$^{300}_{\rm{IS}}$    & $\times$    &   --               &        71.93 &    
   11.14 &        61.82 &         5.07 &     56.7     &        --   &         --
  &          --   & --   &     --  & 0.90  & 49.5 & -- &  0.59 \\    
\hline
\hline
EO4$^{300}_{\rm{IS}}$    & P           &  0.01             &        70.53 &     
  13.85 &        14.87 &         6.86 &     7.93     &        41.81 &         --
  &         5.02 & 8.33         &         2.24 & 1.47  & 48.5 & -- & 0.69  \\   
 
EO7$^{300}_{\rm{IS}}$    & P           &  0.01             &        66.56 &     
  15.68 &        17.02 &         6.50 &     10.4     &        33.86 &         --
  &         5.34 & 6.34         &         2.10  & 1.07 & 45.8 & -- & 0.58  \\   
 
FO4$^{300}_{\rm{IS}}$    & P           &  0.01             &        72.94 &     
  11.75 &        14.15 &         6.23 &     7.85     &        47.04 &         --
  &         4.90 & 9.6          &         2.47  & 1.38 & 50.2 & -- &  0.68 \\   
 
FO7$^{300}_{\rm{IS}}$    & P           &  0.01             &        73.01 &     
  10.83 &        16.13 &         5.07 &     10.9     &        46.06 &         
--   &         4.98 & 9.25         &         2.54 & 0.89 & 50.2 & -- &  0.55\\  
 
\hline 
EO4$^{300}_{\rm{IS}}$    & P           &  0.1              &        70.53 &     
  13.89 &         8.23 &         6.90 &     1.24     &        48.41 &         
--   &         4.36 & 11.10        &         1.99  &1.54 & 48.5 & -- &  0.68 \\ 
  
EO7$^{300}_{\rm{IS}}$    & P           &  0.1              &        66.56 &     
  15.77 &         7.82 &         6.45 &     1.31     &        42.96 &         
--   &         4.42 & 9.72         &         1.66  & 1.04 & 45.8 & -- &  0.59 \\
   
FO4$^{300}_{\rm{IS}}$    & P           &  0.1              &        72.94 &     
  11.71 &         7.73 &         6.40 &     1.22     &        53.50 &         
--   &         4.25 & 12.59        &         2.12  & 1.50 & 50.2 & -- &  0.68 \\
   
FO7$^{300}_{\rm{IS}}$    & P           &  0.1              &        73.01 &     
  10.87 &         6.86 &         5.30 &     1.47     &        55.29 &         
--   &         4.19 & 13.2         &         2.03  & 0.92 & 50.2 & -- &  0.58 \\
   
\hline
\hline
EO4$^{300}_{\rm{IS}}$    &  A          &  0.01             &        70.53 &     
  13.83 &        15.98 &         6.95 &     8.97     &        50.89 &       
10.17 &         5.12 & 9.94         &         3.08  & 1.56 & 48.5 & 11.5 & 0.69 
\\     
EO7$^{300}_{\rm{IS}}$    &  A          &  0.01             &        66.56 &     
  15.71 &        18.53 &         6.22 &     12.3     &        40.38 &        
8.07 &         5.47 & 7.38         &         2.67  & 0.99 & 45.8 & 10.2 & 0.58 
\\     
FO4$^{300}_{\rm{IS}}$    &  A          &  0.01             &        72.94 &     
  11.79 &        15.06 &         6.27 &     8.73     &        57.60 &       
11.51 &         4.94 & 11.66        &         3.21 & 1.44 & 50.2 & 11.9 &0.68 \\
    
FO7$^{300}_{\rm{IS}}$    &  A          &  0.01             &        73.01 &     
  10.86 &        17.79 &         5.36 &     12.4     &        55.45 &       
11.08 &         5.07 & 10.94        &         3.09  & 0.94 & 50.2 & 11.8 & 0.57
\\    
\hline  
EO4$^{300}_{\rm{IS}}$    &  A          &  0.1              &        70.53 &     
  13.95 &         8.55 &         7.05 &     1.42     &        60.03 &       
12.00 &         4.43 & 13.55        &         2.51  & 1.65 & 48.5 & 9.83 & 0.69
\\     
EO7$^{300}_{\rm{IS}}$    &  A          &  0.1              &        66.56 &     
  15.89 &         8.23 &         6.58 &     1.58     &        53.03 &       
10.60 &         4.50 & 11.78        &         2.07  & 1.13 & 45.8 & 7.76 & 0.58
\\     
FO4$^{300}_{\rm{IS}}$    &  A          &  0.1              &        72.94 &     
  11.89 &         8.04 &         6.61 &     1.36     &        66.25 &       
13.24 &         4.29 & 15.44        &         2.65  & 1.65 & 50.2 & 10.1 & 0.70
\\     
FO7$^{300}_{\rm{IS}}$    &  A          &  0.1              &        73.01 &     
  11.04 &         7.40 &         5.72 &     1.59     &        68.21 &       
13.63 &         4.23 & 16.13        &         2.49  & 1.09 & 50.2 & 9.42 & 0.57
\\     
\bottomrule
\end{tabular} %}
\flushleft
\parbox{0.9\linewidth}{\footnotesize 
\textit{Notes.} (1) Model name. $(2)$ Star formation scheme adopted
($\times$=noSF, P=pSF, A=aSF).
$(3)$ Star formation efficiency.
$(4)-(5)$ Cumulative mass injected by the evolution of the original stellar
population, and mass
escaped from the numerical grid, respectively.
Differences in $M_\mathrm{inj}$ for models of same $L_B$ are accounted for by different
sampling of the stellar density profile over the numerical grid.
$(6)$ Total ISM mass retained within the galaxy.
$(7)-(8)$ ISM mass with $T>10^{6}$ K, and $T< 2\times 10^4$~K, respectively.
$(9)$ Mass of the new stars. $(10)$ Cumulative mass ejected by type II SNe. 
$(11)-(13)$ Mean formation time of the new stars, mean SFR, and SFR at the end
of the simulations.
$(14)$ ISM X-ray luminosity in the 0.3--8 keV band.
$(15)-(16)$: SNIa and SNII kinetic energy input per unit time (Sect. 2.2.2).
(17) ISM X-ray emission weighted temperature in the 0.3--8 keV band.
}
\label{tab2}
\end{sidewaystable*}
%\end{table*}

%%%%%%%%%%%%%%%%%%%%%%%%%%%%%%%%%%%%%%%%%%%%%%%%%%%%%%%%%%%%%%%%%%%%%%%%%%%%%%%%
%%%%%%%%%

\renewcommand\arraystretch{1.3}
\setlength{\tabcolsep}{5pt}
\begin{sidewaystable*}
%\begin{table*}
%\centering
\vskip 15truecm
\caption{$\,\,\,\,$ Relevant quantities at $t=13$Gyr for the rotating models
($k=1$) with $\sigma_{\rm e8}$=200 km s$^{-1}$.} 
%\scriptsize
\small
\begin{tabular}{ccccccccccccccccc}
\toprule
name                    &     SF      & $\eta_\mathrm{SF}$&     $M_\mathrm{inj}$  &   
$M_\mathrm{esc}$   &$M_\mathrm{gas}$    &    $M_\mathrm{hot}$   &   $M_\mathrm{c}$&       $M_*^{\rm
new}$&        
$M_\mathrm{inj}^{\rm II}$&  $\langle t\rangle _*^{\rm new}$ & 
$M_*^{\rm new}/\langle t\rangle _*^{\rm new}$ &        $\mathrm{SFR}$ & $\Lx$ & $L_{\rm
SNIa}$ & $L_{\rm SNII}$ & $\Tx$ \\
                       &             &                   & $(10^9$M$_{\sun})$   
 &  $(10^9$M$_{\sun})$     & $(10^9$M$_{\sun})$    & $(10^9$M$_{\sun})$   &  
$(10^9$M$_{\sun})$    & $(10^9$M$_{\sun})$  &  $(10^9$M$_{\sun})$    & (Gyr)    
   & (M$_{\sun}$ yr$^{-1}$)  & (M$_{\sun}$ yr$^{-1}$)   & $(10^{40}$erg
s$^{-1})$ & $(10^{40}$erg s$^{-1})$  & $(10^{40}$erg s$^{-1})$  & (keV) \\
(1)                     &   (2)       &          (3)      &    (4)       &    
(5)      &  (6)         &     (7)      &    (8)       &   (9)        &       
(10)  &  (11)        & (12)         & (13)     & (14) & (15) & (16) & (17)    
\\
\midrule

EO4$^{200}_{\rm{IS}}$    & $\times$    &   -               &        11.92 &     
   9.76 &         2.39 &         0.15 &     2.24     &          -   &          -
  &          -   &         --   &          -   & 0.0007 & 10.2 & -- & 0.50 \\
EO7$^{200}_{\rm{IS}}$    & $\times$    &   -               &        11.93 &     
  10.34 &         1.84 &         0.24 &     1.57     &          -   &          -
  &          -   &         --   &          -  & 0.002 & 10.2 & -- &  0.49 \\
FO4$^{200}_{\rm{IS}}$    & $\times$    &   -               &        12.00 &     
   3.32 &         8.80 &         1.73 &     6.97     &          -   &          -
  &          -   &         --   &          -   & 0.449 & 10.3 & -- & 0.37\\
FO7$^{200}_{\rm{IS}}$    & $\times$    &   -               &        11.96 &     
   3.41 &         8.69 &         1.36 &     7.16     &          -   &          -
  &          -   &         --   &          -   & 0.248 & 10.2 & -- & 0.32\\
 \hline
\hline
EO4$^{200}_{\rm{IS}}$    &  P          &  0.01             &        12.07 &     
   4.32 &         3.28 &         1.63 &     1.57     &         4.47 &          -
  &         6.61 &         0.68 &         0.45 & 0.110 & 10.3 & -- & 0.39 \\
EO7$^{200}_{\rm{IS}}$    &  P          &  0.01             &        12.09 &     
  10.43 &         0.34 &         0.27 &     3.99E-2  &         1.32 &          -
  &         3.84 &         0.34 &         0.02 & 0.003& 10.3 & -- & 0.50 \\
FO4$^{200}_{\rm{IS}}$    &  P          &  0.01             &        12.14 &     
   6.65 &         0.68 &         0.43 &     2.19E-1  &         4.80 &          -
  &         4.20 &         1.14 &         0.08 & 0.006& 10.4 & -- & 0.40 \\
FO7$^{200}_{\rm{IS}}$    &  P          &  0.01             &        12.08 &     
   2.98 &         3.12 &         1.57 &     1.40     &         5.99 &          -
  &         5.39 &         1.11 &         0.37 & 0.163 & 10.3 & -- & 0.36 \\
 \hline
EO4$^{200}_{\rm{IS}}$    &  P          &  0.1              &        12.07 &     
   4.65 &         2.00 &         1.77 &     7.31E-2  &         5.42 &          -
  &         5.74 &         0.94 &         0.12 & 0.086 & 10.3 & -- & 0.40\\
EO7$^{200}_{\rm{IS}}$    &  P          &  0.1              &        12.09 &     
  10.07 &         0.32 &         0.29 &     0.00     &         1.71 &          -
  &         2.37 &         0.72 &         0.00  & 0.003 & 10.3 & -- & 0.51 \\
FO4$^{200}_{\rm{IS}}$    &  P          &  0.1              &        12.14 &     
   3.75 &         3.64 &         2.88 &     8.98E-4  &         4.74 &          -
  &         3.28 &         1.45 &         0.08 & 0.587 & 10.4 & -- & 0.43 \\
FO7$^{200}_{\rm{IS}}$    &  P          &  0.1              &        12.08 &     
   5.79 &         0.49 &         0.40 &     1.98E-2  &         5.81 &          -
  &         3.43 &         1.69 &         0.04 & 0.010 & 10.3 & -- & 0.50 \\
\hline
\hline
EO4$^{200}_{\rm{IS}}$    &  A          &  0.01             &        12.07 &     
   9.40 &         0.27 &         0.17 &     9.92E-2  &         3.00 &        
0.60 &         3.90 &         0.77 &         0.05 & 0.001 & 10.3 & 0.18 & 0.54
\\
EO7$^{200}_{\rm{IS}}$    &  A          &  0.01             &        12.09 &     
   9.70 &         0.47 &         0.24 &     2.09E-1  &         2.40 &        
0.48 &         4.39 &         0.55 &         0.06 & 0.002 & 10.32 & 0.24 & 0.49
\\
FO4$^{200}_{\rm{IS}}$    &  A          &  0.01             &        12.14 &     
   6.77 &         0.98 &         0.72 &     1.86E-1  &         5.47 &        
1.09 &         4.11 &         1.33 &         0.09 & 0.011 & 10.4 & 0.36 & 0.43
\\
FO7$^{200}_{\rm{IS}}$    &  A          &  0.01             &        12.08 &     
   3.09 &         3.77 &         1.42 &     2.19     &         6.53 &        
1.30 &         5.27 &         1.24 &         0.69 & 0.460 & 10.3 & 2.58 & 0.29
\\
 \hline
EO4$^{200}_{\rm{IS}}$    &  A          &  0.1              &        12.07 &     
   7.77 &         0.29 &         0.28 &     7.34E-4  &         5.02 &        
1.00 &         3.72 &         1.35 &         0.00 & 0.002 & 10.3 & 0.01 & 0.53\\
EO7$^{200}_{\rm{IS}}$    &  A          &  0.1              &        12.09 &     
   8.27 &         0.80 &         0.62 &     5.66E-2  &         3.78 &        
0.76 &         5.34 &         0.71 &         0.06 & 0.012 & 10.3 & 0.24 & 0.44
\\
FO4$^{200}_{\rm{IS}}$    &  A          &  0.1              &        12.14 &     
   6.00 &         1.48 &         1.19 &     8.70E-4  &         5.81 &        
1.16 &         3.10 &         1.88 &         0.01 & 0.041 & 10.4 & 0.05 & 0.44\\
FO7$^{200}_{\rm{IS}}$    &  A          &  0.1              &        12.08 &     
   3.82 &         3.82 &         3.28 &     2.55E-3  &         5.57 &        
1.11 &         2.92 &         1.91 &         0.11 & 0.982 & 10.3 & 0.40 & 0.41
\\
\bottomrule
\end{tabular} %}
\flushleft
\parbox{0.9\linewidth}{\footnotesize 
\textit{Notes.} All quantities as in Tab. \ref{tab2}.
}
\label{tab3}
\end{sidewaystable*}

\vfill
\eject
\clearpage

\renewcommand\arraystretch{1.3}
\setlength{\tabcolsep}{5pt}
\begin{sidewaystable*}
%\begin{table*}
%\centering
\vskip 15truecm
\caption{$\,\,\,\,$ Relevant quantities at $t=13$Gyr for the rotating
  models ($k=1$) with $\sigma_{\rm e8}$=250 km s$^{-1}$ (aSF only).} 
%\scriptsize
\small
\begin{tabular}{ccccccccccccccccc}
\toprule
name                    &     SF      & $\eta_\mathrm{SF}$&     $M_\mathrm{inj}$  &   
$M_\mathrm{esc}$   &$M_\mathrm{gas}$    &    $M_\mathrm{hot}$   &   $M_\mathrm{c}$&       $M_*^{\rm
new}$&        $M_\mathrm{inj}^{\rm II}$&  $\langle t\rangle _*^{\rm new}$ & 
$M_*^{\rm new}/\langle t\rangle _*^{\rm new}$ &        $\mathrm{SFR}$ & $\Lx$ & $L_{\rm
SNIa}$ & $L_{\rm SNII}$ & $\Tx$ \\
                       &             &                   & $(10^9$M$_{\sun})$   
 &  $(10^9$M$_{\sun})$     & $(10^9$M$_{\sun})$    & $(10^9$M$_{\sun})$   &  
$(10^9$M$_{\sun})$    & $(10^9$M$_{\sun})$  &  $(10^9$M$_{\sun})$    & (Gyr)    
   & (M$_{\sun}$ yr$^{-1}$)  & (M$_{\sun}$ yr$^{-1}$)   & $(10^{40}$erg
s$^{-1})$ & $(10^{40}$erg s$^{-1})$  & $(10^{40}$erg s$^{-1})$  & (keV) \\
(1)                     &   (2)       &          (3)      &    (4)       &    
(5)      &  (6)         &     (7)      &    (8)       &   (9)        &       
(10)  &  (11)        & (12)         & (13)     & (14) & (15) & (16) & (17)    
\\
\midrule

EO4$^{250}_{\rm{IS}}$    & $\times$    &   --              &        31.68 &     
   8.28 &        23.84 &         4.02 &     19.8 &          --  &          --  &
         --  &          --  &          --  & 0.76 & 23.9 & -- & 0.55 \\
EO7$^{250}_{\rm{IS}}$    & $\times$    &   --              &        30.63 &     
  11.42 &        19.71 &         3.42 &     16.3 &          --  &          --  &
         --  &          --  &          --  & 0.33 & 23.1 & -- & 0.55 \\
\hline
\hline
EO4$^{250}_{\rm{IS}}$    &  A          &  0.1              &        32.14 &     
   8.39 &         4.84 &         4.41 &     0.40 &        23.64 &         4.72 &
        4.35 &         5.43 &         0.86 & 1.01 & 24.2 & 3.25 & 0.57 \\
EO7$^{250}_{\rm{IS}}$    &  A          &  0.1              &        31.10 &     
  11.42 &         4.47 &         4.31 &     0.16 &        19.01 &         3.80 &
        4.13 &         4.60 &         0.42 & 0.66 & 23.4 & 1.57 & 0.58 \\
\midrule                                      
EO4$^{250}_{\rm{IS}}$    &  A          &  0.01             &        32.14 &     
   8.28 &         7.17 &         4.06 &     3.08 &        20.87 &         4.17 &
        5.06 &         4.12 &         1.14 & 0.82 & 24.2 & 4.31 & 0.56 \\
EO7$^{250}_{\rm{IS}}$    &  A          &  0.01             &        31.10 &     
  11.18 &         6.92 &         4.00 &     2.90 &        16.24 &         3.24 &
        5.32 &         3.05 &         0.93 & 0.51 & 23.4 & 3.55 & 0.52 \\
\bottomrule
\end{tabular}
\flushleft
\parbox{0.9\linewidth}{\footnotesize %
\textit{Notes}. All quantites are as in Tab. \ref{tab2}.}
\label{tab4}
\end{sidewaystable*}
%\end{table*}

\label{LastPage}
\renewcommand\arraystretch{1.3}
\setlength{\tabcolsep}{5pt}
\begin{sidewaystable*}
%\begin{table*}
%\centering
\vskip 15truecm
\caption{$\,\,\,\,$ Relevant quantities at $t=13$Gyr for the mildly
  rotating models ($k=0.1$) with $\sigma_{\rm e8}$=250 km s$^{-1}$ (aSF only).} 
%\scriptsize
\small
\begin{tabular}{ccccccccccccccccc}
\toprule
name                    &     SF      & $\eta_\mathrm{SF}$&     $M_\mathrm{inj}$  &   
$M_\mathrm{esc}$   &$M_\mathrm{gas}$    &    $M_\mathrm{hot}$   &   $M_\mathrm{c}$&       $M_*^{\rm
new}$&        $M_\mathrm{inj}^{\rm II}$&  $\langle t\rangle _*^{\rm new}$ & 
$M_*^{\rm new}/\langle t\rangle _*^{\rm new}$ &        $\mathrm{SFR}$ & $\Lx$ & $L_{\rm
SNIa}$ & $L_{\rm SNII}$ & $\Tx$ \\
                       &             &                   & $(10^9$M$_{\sun})$   
 &  $(10^9$M$_{\sun})$     & $(10^9$M$_{\sun})$    & $(10^9$M$_{\sun})$   &  
$(10^9$M$_{\sun})$    & $(10^9$M$_{\sun})$  &  $(10^9$M$_{\sun})$    & (Gyr)    
   & (M$_{\sun}$ yr$^{-1}$)  & (M$_{\sun}$ yr$^{-1}$)   & $(10^{40}$erg
s$^{-1})$ & $(10^{40}$erg s$^{-1})$  & $(10^{40}$erg s$^{-1})$  & (keV) \\
(1)                     &   (2)       &          (3)      &    (4)       &    
(5)      &  (6)         &     (7)      &    (8)       &   (9)        &       
(10)  &  (11)        & (12)         & (13)     & (14) & (15) & (16) & (17)    
\\
\midrule

EO4$^{250}_{k=0.1  }$    & $\times$    &   --              &        32.14 &     
   8.63 &        23.51 &         4.78 &     18.7 &         --   &         --   &
         --  &          --  &         --   & 5.82 & 24.2 & -- & 0.69 \\
EO7$^{250}_{k=0.1  }$    & $\times$    &   --              &        31.10 &     
  12.94 &        18.17 &         3.44 &     14.7 &         --   &         --   &
         --  &          --  &         --   & 2.20 & 23.4 & -- & 0.69 \\
\hline
\hline
EO4$^{250}_{k\rm{=0.1}}$    &  A          &  0.1              &        32.14 &  
      8.46 &         5.13 &         5.04 &     0.086 &        23.20 &        
4.64 &         4.28 &         5.42 &         1.01 & 6.33 & 24.2 & 3.76 & 0.67 
\\
EO7$^{250}_{k\rm{=0.1}}$    &  A          &  0.1              &        31.10 &  
     13.06 &         3.63 &         3.56 &    0.065 &        18.01 &        
3.60 &         4.03 &         4.47 &         0.45 & 2.78 & 23.4 & 1.76 & 0.68 \\
 \hline                                  
EO4$^{250}_{k\rm{=0.1}}$    &  A          &  0.01             &        32.14 &  
      8.58 &         5.28 &         4.84 &     0.442 &        22.84 &        
4.56 &         4.46 &         5.12 &        0.91 & 6.39 & 24.2 & 3.69 & 0.66\\
EO7$^{250}_{k\rm{=0.1}}$    &  A          &  0.01             &        31.10 &  
     12.96 &         4.14 &         3.48 &    0.656 &        17.50 &        
3.50 &         4.22 &         4.14 &        0.63 & 2.51 & 23.4 & 2.32 & 0.68 \\
\bottomrule
\end{tabular}
\flushleft
\parbox{0.9\linewidth}{\footnotesize %
\textit{Notes}. All quantities are as in Tab.~\ref{tab2}.}
\label{tab5}
\end{sidewaystable*}
%\end{table*}

\end{document}